\documentclass[aps,prb,twocolumn,showpacs,amsmath,floatfix,intlimits]{revtex4-1}

\usepackage{amsfonts}
\usepackage{graphicx}
\usepackage{braket}
\usepackage{overbar}
\usepackage{mathtools}
\usepackage[normalem]{ulem}
\usepackage{color,soul} 

\newcommand{\MoSe}[1]{MoSe\ensuremath{{}_{#1}}}
\newcommand{\MoS}[1]{MoS\ensuremath{{}_{#1}}}
\newcommand{\grid}[1]{\ensuremath{{#1}\times{#1}\times{1}}}
\newcommand{\Grid}[1]{\ensuremath{{#1}\times{#1}\times{#1}}}
\newcommand{\angs}[1]{\ensuremath{{#1}\,\textrm{\AA}}}
\newcommand{\Gperp}{\ensuremath{\mathbf{G}_{\perp}}}

\newcommand{\Qo}{{\ensuremath{\mathbf{q}{\rightarrow}0}}}
\newcommand{\vorocell}[1]{{\ensuremath{\mathcal{C}_\mathbf{#1}}}}

\begin{document}

\title{Non-uniform sampling schemes of the Brillouin zone for many-electron perturbation-theory calculations in reduced dimensionality}

\author{Felipe H. da Jornada}
\thanks{These two authors contributed equally.}
\affiliation{Department of Physics, University of California at
Berkeley, California 94720}
\affiliation{Materials Sciences Division, Lawrence Berkeley
National Laboratory, Berkeley, California 94720}

\author{Diana Y. Qiu}
\thanks{These two authors contributed equally.}
\affiliation{Department of Physics, University of California at
Berkeley, California 94720}
\affiliation{Materials Sciences Division, Lawrence Berkeley
National Laboratory, Berkeley, California 94720}

\author{Steven G. Louie}
\email[Email: ]{sglouie@berkeley.edu}
\affiliation{Department of Physics, University of California at
Berkeley, California 94720}
\affiliation{Materials Sciences Division, Lawrence Berkeley
National Laboratory, Berkeley, California 94720}

\date{\today}

\begin{abstract}
First principles calculations based on many-electron perturbation theory methods, such as the \textit{ab initio} GW and GW plus Bethe-Salpeter equation (GW-BSE) approach, are reliable ways to predict quasiparticle and optical properties of materials, respectively. However, these methods involve more care in treating the electron-electron interaction and are considerably more computationally demanding when applied to systems with reduced dimensionality, since the electronic confinement leads a slower convergence of sums over the Brillouin zone due to a much more complicated screening environment that manifests in the "head" and "neck" elements of the dielectric matrix. Here, we present two new schemes to sample the Brillouin zone for GW and GW-BSE calculations: the non-uniform neck subsampling method and the clustered sampling interpolation method, which can respectively be used for a family of single-particle problems, such as GW calculations, and for problems involving the scattering of two-particle states, such as when solving the BSE. We tested these methods on several few-layer semiconductors and graphene and show that they perform a much more efficient sampling of the Brillouin zone and yield two to three orders of magnitude reduction in the computer time. These two methods can be readily incorporated into several \textit{ab initio} packages that compute electronic and optical properties through the GW and GW-BSE approaches.
\end{abstract}

\pacs{73.22.-f, 71.35.-y, 78.67.-n}

\maketitle

\section{Introduction}

Many-electron perturbation theory methods, especially those based on density-functional theory (DFT) as the starting mean field, are becoming increasingly popular for predicting electronic excited-state properties of novel materials. Some of the most commonly used methods of this family include: the \textit{ab initio} GW approximation, which allows for the computation of 
quasiparticle (QP) properties of materials~\cite{hedin65,hybertsen86}; the GW plus Bethe-Salpeter equation (GW-BSE) method, which accesses correlated two-particle states such as excitons~\cite{strinati88,rohlfing00}; and the adiabatic-connection fluctuation-dissipation theorem (ACFDT) methods, which allow for accurate computation of the total energy of materials~\cite{Fuchs02,Marini06,Lu09,Toulouse09,Lebegue10}, among others. These methods are now available in a variety of mature and optimized computer packages~\cite{jdeslip11BGW,YAMBO,ABINIT,GPAW} and have been applied with success to predict electronic and optical properties of a variety of different systems, ranging from bulk 3D semiconductors to systems with reduced dimensionality, such as molecules, graphene, carbon nanotubes, and nanoribbons.

More recently, there has been interest in applying this family of methods to quasi-two-dimensional (quasi-2D) semiconducting materials, which was motivated by the experimental isolation of monolayer transition metal dichalcogenides (TMDs) such as \MoS2 and \MoSe2. However, even though the conceptual approximations employed on conventional 3D systems still hold for quasi-2D materials, it has been notoriously harder to perform \textit{ab initio} many-electron perturbation theory calculations on these monolayer TMDs. For example, while one can typically converge GW QP energies on bulk Si with a \Grid{4} $k$ grid, one needs a much finer $k$ grid of \grid{24} to converge the quasiparticle gap of monolayer \MoS2~\cite{qiu13,wirtz13,huser13,qiu16,rasmussen16}. This is unexpected at first, since: (1) the ground state properties of monolayer \MoS2 calculated with DFT within the local density approximation (LDA) converge on a much coarser $\sim\grid{6}$ $k$ grid; and (2) monolayer \MoS2 has a larger bandgap than Si, so one might naively expect that a coarser $k$ grid is enough to converge the electronic properties of monolayer \MoS2.

The difficulty in converging the electronic properties of quasi-2D semiconductors with $k$-point sampling is an indirect manifestation of unusual features in electron-electron interactions in these systems. In a plane wave basis set, these features are encoded in the dielectric matrix $\varepsilon_{\mathbf{G, G'}}(\mathbf{q})$, which displays a strong, sharply-peaked feature in its $\mathbf{q}$ dispersion in the long wavelength limit not found in typical bulk semiconductors~\cite{beigi06,cudazzo11,huser13,chernikov14,qiu16,rasmussen16}. These features in the dielectric screening manifest in a small portion of the Brillouin zone when performing many-electron perturbation theory calculations, and give rise to the very slow convergence with respect to the number of $q$-vectors included when computing the GW quasiparticle self energy of systems with reduced dimensionality.

In this paper, we address this issue of slow convergence of many-electron perturbation theory calculations with $q$-point sampling. We introduce two new methods here, the non-uniform neck subsampling (NNS) method, which provides an efficient way to sample the Brillouin zone and capture features of the dielectric matrix due the electronic confinement, which can be readily used in GW and ACFDT calculations; and the clustered sampling interpolation (CSI) method, which is an approximation to efficiently compute matrix elements which arise in two-body problems, such as in the context of solving the Bethe-Salpeter equation. Specifically for the case of calculating the self-energy and excitonic effects on mono- or few-layer transition metal dichalcogenides, we show that these methods perform a much more efficient sampling of the Brillouin zone and yields orders of magnitude reduction in the computer run time. Moreover, our methods do not assume any analytical form of the dielectric screening~\cite{beigi06,rasmussen16}, and can be equally applied to 1D and 2D semiconducting and metallic systems. These two methods can be readily incorporated into several \textit{ab initio} packages that compute electronic and optical properties through many-electron perturbation theory methods.

Our paper is organized as follows: in Section~\ref{sec:subsampling}, we introduce the non-uniform neck subsampling (NNS) method to efficiently calculate sums in the Brillouin zone involving the screened Coulomb interaction. Our main results are in Eqs.~\ref{eqn:subsampling1} and~\ref{eqn:subsampling2}, and the NNS is summarized graphically in Fig.~\ref{fig:diagram}. We give example of the NNS method by performing calculations on bilayer \MoSe2 and graphene. In Section~\ref{sec:clustered}, we develop the cluster sampling interpolation (CSI) method. The main result of this part is Eq.~\ref{eqn:csi}, and the speed up due to the method is presented in Figs.~\ref{fig:csiconverge} and~\ref{fig:csiP}. We conclude in Section~\ref{sec:conclusion} by summarizing our results.

\section{\label{sec:subsampling} Non-uniform neck subsampling (NNS) method}

In this section, we introduce a method to perform non-uniform sampling of the Brillouin zone. Our goal is to efficiently evaluate sums that are common in many-electron perturbation theory calculations with plane-wave basis sets, which involve the screened Coulomb interaction matrix $W_\mathbf{G G'}(\mathbf{q}, \omega)$. In general, we will be interested in evaluating sums over the Brillouin zone with the form
\begin{equation}
\label{eqn:general}
I_{\mathbf{G G'}}(\omega) = \sum_\mathbf{q}
	B_{\mathbf{G G'}}(\mathbf{q}, \omega) \,
	W_\mathbf{G G'}\left( \mathbf{q}, \omega \right),
\end{equation}
where $\mathbf{q}$ is a transferred momentum, or $q$-vector, typically defined on a uniform, $\Gamma$-centered Monkhorst-Pack grid, $\mathbf{G}$ and $\mathbf{G'}$ are reciprocal lattice vectors, $B_{\mathbf{G G'}}$ is a smooth function, and the screened Coulomb interaction is $W_\mathbf{G G'}(\mathbf{q}, \omega) = \varepsilon^{-1}_\mathbf{G G'}(\mathbf{q}, \omega) \, v(\mathbf{q+G'})$, where $\varepsilon^{-1}(\mathbf{q})$ is the dielectric matrix and $v(\mathbf{q})$ is the bare Coulomb interaction.

We will be particularly interested in evaluating sums related to the electronic self energy, such as the screened-exchange contribution to the GW self energy,
\begin{align}
\label{eqn:SX}
\Sigma^\mathrm{sx}_{n\mathbf{k}}(\omega) &= -
	\sum_{v \mathbf{G G'}} \left[ \sum_{\mathbf{q}}
	 B^\mathrm{sx} (\mathbf{q}) \,
	 	W_\mathbf{G G'}(\mathbf{q}, \omega - E_{v \mathbf{k-q}}) \right] \notag \\
B^\mathrm{sx} (\mathbf{q}) &=
\braket{u_{n \mathbf{k}} | e^{i\mathbf{G\cdot r}} | u_{v \mathbf{k-q}} }
\braket{u_{v \mathbf{k-q}} | e^{-i\mathbf{G'\cdot r}} | u_{n \mathbf{k}} } ,
\end{align}
where $v$ denotes an occupied band and the indices $n, \mathbf{k, G,}$ and $\mathbf{G'}$ are implicit in $B^\mathrm{sx} (\mathbf{q})$.

Eq.~\ref{eqn:general} cannot be applied directly on semiconductors because of the divergence of the Coulomb interaction at $\mathbf{q}{=}0$. While several treatments have been proposed to deal with this divergence~\cite{gygi86,carrier07}, we restrict our discussion to one particular stochastic method which is well-suited for systems with reduced dimensionality. We start by taking the continuous limit and then re-discretizing Eq.~\ref{eqn:general}, assuming the matrix elements $B(\mathbf{q})$ to be smooth. This yields
\begin{align}
\label{eqn:MCavg}
I_{\mathbf{G G'}}(\omega) =& \sum_{\mathbf{q}}
	B_{\mathbf{G G'}}(\mathbf{q}, \omega) \,
	\widebar{W}_\mathbf{G G'}\left( \mathbf{q}, \omega \right), \\
\label{eqn:Wavg}
\widebar{W}_\mathbf{G G'}\left(\mathbf{q}, \omega \right) =&
	\frac{1}{V_\mathbf{q}} \int_{\vorocell{q}} \mathrm{d}^Dq' \;
	W_\mathbf{G G'}\left( \mathbf{q}', \omega \right),
\end{align}
    where $D$ denotes the number of dimensions on the system. Each integral is performed over the Voronoi cell that surrounds each $\mathbf{q}$ vector, denoted by \vorocell{q}~\footnote{A Voronoi cell can be thought of as a generalization of the Wigner-Seitz cell for non-uniform grids; if the $q$-vectors are sampled uniformly on a $\Grid{n_q}$ grid, as is usually the case for 3D systems, then \vorocell{q} is simply the Brillouin zone scaled down isotropically by $n_q$.}.We denote the volume (or area/length for 2D/1D systems) of \vorocell{q} by $V_\mathbf{q}$.

To evaluate the integral in Eq.~\ref{eqn:Wavg}, it is possible to employ the exact analytic behavior of $\varepsilon^{-1}_\mathbf{G G'}(\Qo)$ and use a Monte Carlo average scheme to more efficiently sample Brillouin zones with arbitrary shapes. This stochastic approach is used in a few GW packages~\cite{jdeslip11BGW,YAMBO}. For example, for bulk 3D semiconductors, $\varepsilon^{-1}_\mathbf{G G'}(\Qo)$ is smooth and approaches a constant as \Qo{}, so the integral in Eq.~\ref{eqn:Wavg} can be evaluated for 3D semiconductors as
\begin{align}
\label{eqn:Wavg_MC_3d}
\begin{split}
\widebar{W}_\mathbf{G G'}^\mathrm{MC}\left( \mathbf{q}{\ne}0, \omega \right) &=
	 \varepsilon^{-1}_\mathbf{G G'}(\mathbf{q}, \omega)\;
	 v^\mathrm{MC} (\mathbf{q}, \mathbf{G}'), \\
\widebar{W}_\mathbf{G G'}^\mathrm{MC}\left( \mathbf{q}{=}0, \omega \right) &=
	 \varepsilon^{-1}_\mathbf{G G'}(\mathbf{q}_0, \omega)\;
	 v^\mathrm{MC} (0, \mathbf{G}'), \\
v^\mathrm{MC} (\mathbf{q}, \mathbf{G}) &\coloneqq
	\frac{1}{N_\mathrm{MC}}
	\sum_{\mathbf{q}_\mathrm{MC} \in \vorocell{q}}
	v( \mathbf{q}_\mathrm{MC} + \mathbf{G}),
\end{split}
\end{align}
where a small but finite vector $\mathbf{q}_0$ is employed to compute the long wavelength limit of the dielectric matrix, and a  value of $N_\mathrm{MC}\sim\times10^6$ Monte Carlo samples $\{\mathbf{q}_\mathrm{MC} \in \vorocell{q}\}$ is typically enough to converge the sum to within a few meVs.

Once we truncate the Coulomb potential~\cite{beigi06}, Eqs.~\ref{eqn:MCavg} and~\ref{eqn:Wavg_MC_3d} can also be applied on quasi-2D semiconductors. In fact, it is possible to converge the \emph{absolute} Fock exchange energy on bilayer \MoSe2 to within 70~meV on a \grid{6} $q$ grid, which shows that the matrix elements $B^\mathrm{sx}(\mathbf{q})$ in Eq.~\ref{eqn:SX} are smooth functions even for systems with reduced dimensionality, and that Monte Carlo sampling methods can effectively capture the fast variations in the Coulomb interaction. However, these stochastic methods, as usually employed, become much less efficient to evaluate the total GW self energy for quasi-2D semiconductors. Still for the case of bilayer \MoSe2{}, a \grid{24} $q$ grid is necessary to converge the GW self energy to within 50~meV, even if we use a more sophisticated analytic expression for the  of the inverse dielectric matrix, $\varepsilon_{00}^{-1}(\mathbf{q})$.

The slow convergence of Eq.~\ref{eqn:MCavg} on systems with reduced dimensionality is a sign that the analytic models typically employed for the dielectric matrix are no longer accurate in the range of $\mathbf{q}$- and $\mathbf{G}$-vectors we are interested in. This is mainly due to two factors: first, the dielectric matrices of these systems have many features as a function of $\mathbf{q}$ which are hard to model analytically~\cite{cudazzo11,huser13,chernikov14,qiu16}; and second, these Monte Carlo averages should be performed not only for the head element ($\mathbf{G}{=}\mathbf{G}'{=}0$), but also for a series of reciprocal lattice vectors $\Gperp,\Gperp'$ in the confined direction  (e.g., along the direction of the normal vector for a 2D material), which we denote to as the \emph{neck} elements of the matrix.

The physical motivation for focusing on these neck matrix elements of the dielectric matrix, $\varepsilon_\mathbf{\Gperp \Gperp'}(\mathbf{q})$, is that the \Gperp{} vectors in the confined direction become continuous as the simulation supercell grows in the confined direction, and so they become almost as important as the $\mathbf{G}{=}0$ vector. For example, in our calculation on bilayer \MoSe2, the magnitude of the smallest, nonzero reciprocal lattice vector \Gperp{} corresponding to the out-of-plane direction is 5\% of that of the in-plane, primitive reciprocal lattice vector. Consequently, not only will the head element of the screened Coulomb potential $W_{\mathbf{00}}(\mathbf{q}) = \varepsilon^{-1}_{00}(\mathbf{q}) v(\mathbf{q})$ be large, but a series of neck elements $W_{\mathbf{\Gperp, \Gperp'}}(\mathbf{q})$ will also be large, as long as $|\Gperp|$ and $|\Gperp'|$ are smaller than, or of the same order of magnitude as the $q$-vectors in \vorocell{q{=}0}.

\begin{figure}[tp]
   \includegraphics[width=246.0pt]{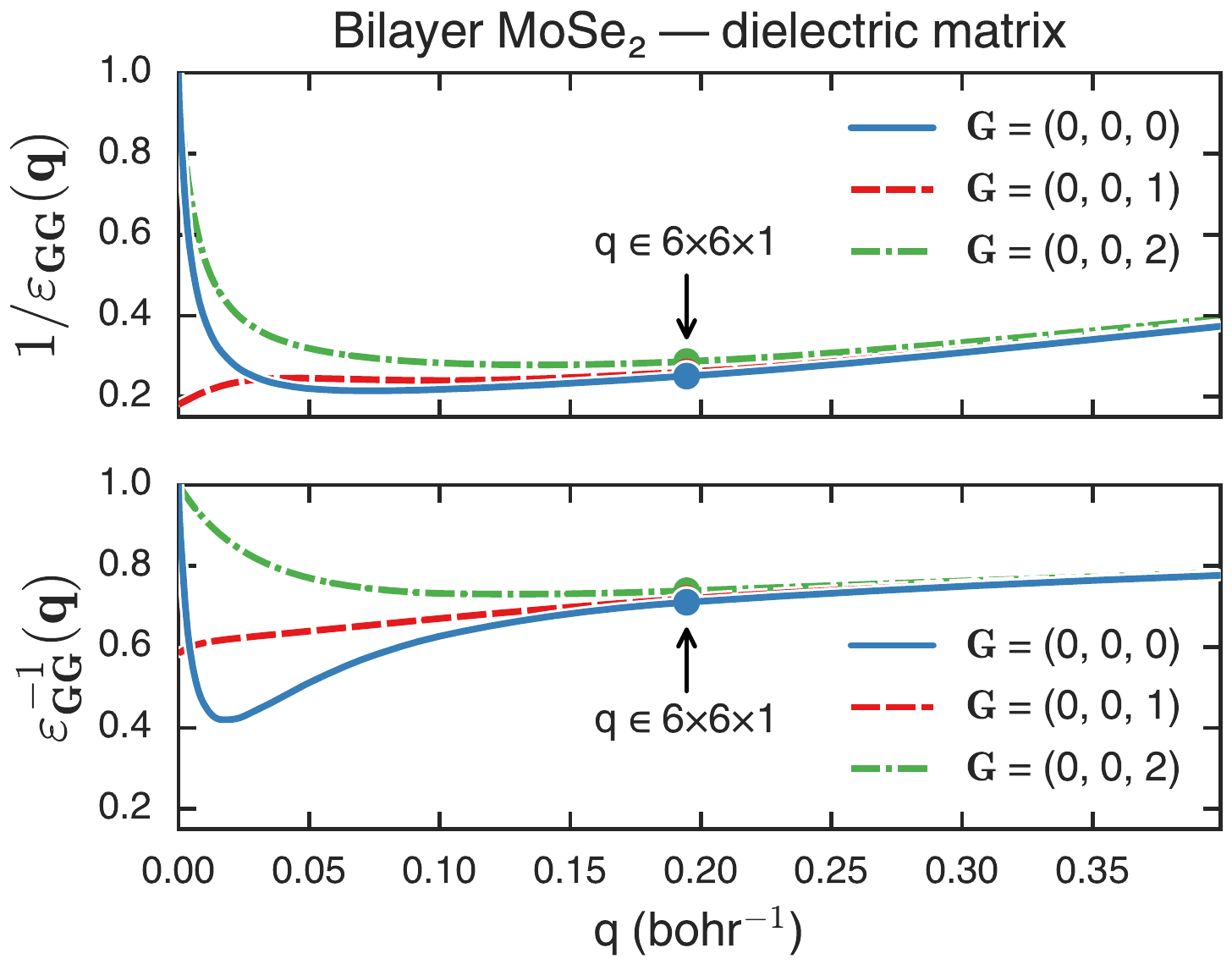}
   \caption{\label{fig:epsinv} (Color online) Selected matrix elements of the dielectric matrix of bilayer \MoSe2: $1/\varepsilon_{\mathbf{G G}}^{-1}(\mathbf{q})$ (top panel) and $\varepsilon_{\mathbf{G G}}^{-1}(\mathbf{q})$ (bottom panel). The difference between the two curves is related to local field effects. Dots represent calculations performed with smallest non-zero $\mathbf{q}$ point from an uniform \grid{6} $q$ grid.}
\end{figure}

We illustrate the sharp features in the inverse dielectric matrix of bilayer \MoSe2 by plotting in Fig.~\ref{fig:epsinv} some selected components of $1/(\varepsilon_{\Gperp, \Gperp})$ and $\varepsilon^{-1}_{\Gperp, \Gperp}$, with $\Gperp=(0,0,G_z)$. It becomes evident that the inverse dielectric matrix has completely different \Qo{} behavior depending on whether $G_z$ is odd or even. More importantly, even for a simple system such as bilayer \MoSe2, local fields play a very important role, as different $\Gperp$ components of the inverse dielectric function $\varepsilon^{-1}_{\Gperp, \Gperp}$ disperse in a qualitatively different way from the matrix elements $1/(\varepsilon_{\Gperp, \Gperp})$. Because these differences are not uniform among different $\Gperp$ components, this shows that local fields mix different components of the neck of the inverse dielectric function in a non-obvious way, in the process of inverting the dielectric matrix.

Since local fields are important in systems with reduced dimensionality, an accurate analytical model for the neck of the inverse dielectric matrix $\varepsilon^{-1}_\mathbf{G, G'}$ requires us to cary out the inversion of the dielectric matrix and explicitly include a series of off-diagonal matrix elements. So, while it is important to capture the $q$-dispersion of $\varepsilon^{-1}_\mathbf{\Gperp \Gperp'}(\mathbf{q})$ in order to evaluate the sums in the Brillouin zone, it seems unlikely that there is a compact and reliable analytic expression for $\varepsilon^{-1}$ for the range of the $\Gperp$- and $q$-vectors we are interested in, especially one that is valid for a wide range of complex materials.

%
Even with no analytical expression for $\varepsilon^{-1}_\mathbf{\Gperp \Gperp'}(\mathbf{q})$, we can still speedup the convergence of the sum in Eq.~\ref{eqn:Wavg} dramatically if we sample more efficiently the inverse dielectric matrix in the region where both the Coulomb interaction is larger and where $\varepsilon^{-1}(\mathbf{q})$ varies the most in \vorocell{0}. We propose to explicitly capture these variations by breaking up the integral in Eq.~\ref{eqn:Wavg} into one radial and one angular part, where the radial part is divided into $N_s$ annuli, each one having a thickness $\Delta_s$. We also approximate the radial integral with a discrete sum over $N_s$ points $q_s$, which we refer to as subsampling points, and write
\begin{align}
\omit\span \widebar{W}_{\Gperp \Gperp'}^\mathrm{sub}( \mathbf{q}{=}0, \omega) \notag\\
\label{eqn:Wavg_radial}
		&\equiv \sum_{s=1}^{N_s} w_s \,
	\varepsilon^{-1}_{\Gperp \Gperp'}(\mathbf{q}_s\, \omega) \, v(\mathbf{q}_s + \Gperp'), \\
\label{eqn:weights}
w_s &\approx \frac{1}{N_\mathrm{MC}} \sum_{\mathbf{q}_\mathrm{MC}}
	\theta(|\mathbf{q}_\mathrm{MC}|-a_s) \theta(a_{s+1}-|\mathbf{q}_\mathrm{MC}|),
\end{align}
where is $w_s$ the weight associated with each subsampling point, and $a_s$ is just a shorthand for the inner radius of each annulus, i.e., $a_s \equiv \sum_{i=1}^{s-1} \Delta_s$.

While the approximation in Eq.~\ref{eqn:Wavg_radial} works best with isotropic materials, we stress that most of the variation of $\varepsilon^{-1}(\mathbf{q})$ only depends on $|\mathbf{q}|$, and we can always choose a direction for each subsampling point $q_s$ that yields the same value for the inverse dielectric function as the angle-averaged inverse dielectric function (at least for one particular pair of G-vectors).

For a given $N_s$ number of subsampled points, we have the freedom to define two quantities: the  $N_s$ subsampling points $q_s$ where the dielectric matrix has to be explicitly computed, and the $N_s$ annulus thicknesses $\Delta_s$. As we derive in the Appendix from simple assumptions of the qualitative behavior of the inverse dielectric matrix, the optimal position of the subsampling point is halfway between between the inner and outer radius of each annulus, $q_s = a_{s} + \frac{\Delta_s}{2}$, where $a_{s+1}=a_s+\Delta_s$.  On the other hand, while the choice of optimal thicknesses is system dependent, a practical solution is to use a polynomial sampling of degree $p$, $\Delta_s = \Delta_1 \times s^p$ (which corresponds to polynomial sampling of degree $p+1$ for the subsampling points). We tested different samplings by calculating the quasiparticle bandgap of bilayer \MoSe2 with a set of $q$-vectors defined on a regular \grid{6} Monkhorst-Pack grid plus a set of $N_s=10$ subsampling $\mathbf{q}$ points. We tested thicknesses generated with a polynomial of degree $p=0$, $p=1$ and $p=2$ and found little difference in the resulting energies, with the \emph{absolute} quasiparticle energies differing by less than 10~meV for states near the  Fermi energy between calculations generated with $p=1$ and $p=2$, and by less than 20~meV between calculations generated with $p=0$ and $p=1$. However, subsampling points generated with a quadratic grid ($p=1$) capture the dip feature in Fig.~\ref{fig:epsinv} better than subsampling points generated with a linear grid ($p=0$), so, for simplicity, we use $p=1$ from here on when performing subsequent calculations.

\begin{figure}[tp]
   \includegraphics[width=246.0pt]{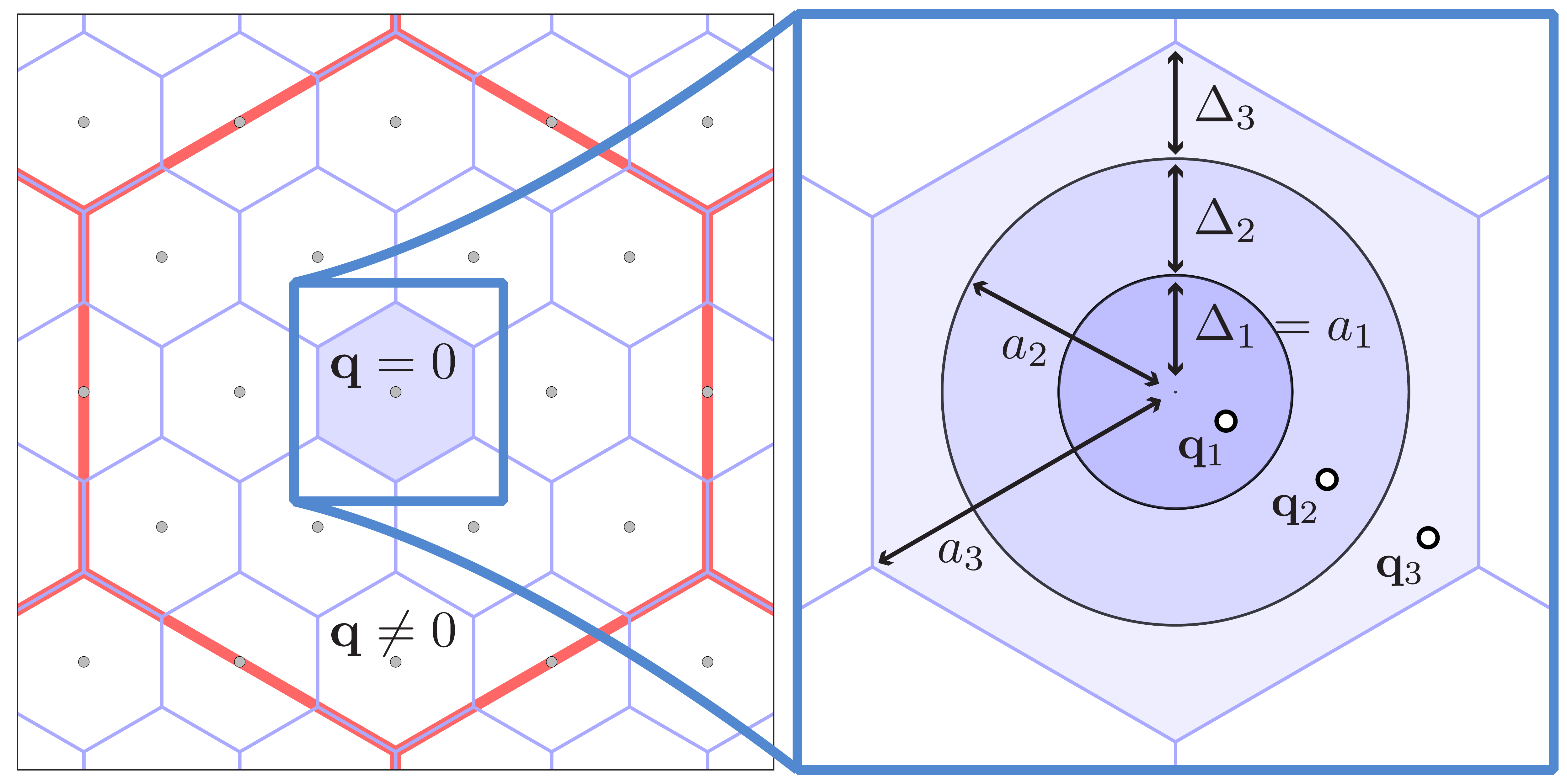}
    \caption{\label{fig:diagram} (Color online) Graphical representation of the NNS scheme. Left panel: $\mathbf{q}$ points involved in the sum in Eq.~\ref{eqn:subsampling1}. Each smaller hexagon represents the Voronoi cell \vorocell{q} that encloses each $\mathbf{q}$ point. The thicker red line denotes the Brillouin zone edge, and each dot on the left panel represents one point where we calculate both the matrix elements $B$ and the screened Coulomb interaction $W$. Right panel: special treatment for $\mathbf{q}{=}0$ point of Eq.~\ref{eqn:subsampling1}. Each dot in this panel represents a $\mathbf{q}$ point where we compute the screened Coulomb interaction. In this example, we use $N_s=3$ subsampled points.}
\end{figure}

In principle, the extra cost associated with the sampling technique would be the ratio of the number of subsampling points to the number of $q$-vectors on the regular grid of $q$-vectors. However, since the fast variations in the dielectric matrix are confined to a small number of $\Gperp$-vectors, we only need to calculate the dielectric matrices for the subsampling contribution in Eq.~\ref{eqn:Wavg_radial} for a small number of neck $\Gperp$-vectors. We choose these vectors on the condition that $|\mathbf{G}|^2 \le E_\mathrm{cut}^\mathrm{sub} \equiv |\mathbf{G}_\parallel^\mathrm{min}|^2$, where $\mathbf{G}_\parallel^\mathrm{min}$ is the smallest reciprocal lattice vector in a periodic direction. The cutoff $E_\mathrm{cut}^\mathrm{sub}$ used for the subsampling points is typically much smaller than the cutoff $E_\mathrm{cut}$ needed for the full dielectric matrix in GW calculations, so the extra cost associated with the NNS scheme is small. For instance, on bilayer \MoSe2, the number of G-vectors up to $E_\mathrm{cut}=35$~Ry  and $E_\mathrm{cut}^\mathrm{sub}\approx1.36$~Ry is 11667 and 37, respectively. We arrive at the final expression to evaluate the sum in Eq.~\ref{eqn:general} within the NNS method,
\begin{widetext}
\begin{align}
\label{eqn:subsampling1}
I_{\mathbf{G G'}}(\omega) &= 
	\sum_\mathbf{q} B_{\mathbf{G G'}}(\mathbf{q}, \omega) \widebar{W}^\mathrm{sub}_\mathbf{G G'}\left( \mathbf{q}, \omega \right) \\
\label{eqn:subsampling2}
\widebar{W}^\mathrm{sub}_{\mathbf{G G'}}(\mathbf{q}{\ne}0, \omega) &= 
		W_{\mathbf{G G'}}(\mathbf{q}, \omega) \notag\\
\widebar{W}^\mathrm{sub}_{\mathbf{G G'}}(\mathbf{q}{=}0, \omega) &= 
\begin{cases}
	W_{\mathbf{G G'}}(\mathbf{q}_0, \omega) &\text{for $|\mathbf{G}|^2$ and $|\mathbf{G}'|^2 > E^\mathrm{sub}_\mathrm{cut}$}\\
	\sum_{s=1}^{N_s} w_s W_{\mathbf{G G'}}(\mathbf{q}_s, \omega) &\text{otherwise}.
\end{cases}
\end{align}
\end{widetext}
where $\mathbf{q}_0$ is an arbitrarily small but non-zero vector. Eqs.~\ref{eqn:subsampling1} and~\ref{eqn:subsampling2}, together with the definition of the subsampling weight in Eq.~\ref{eqn:weights} form the basis of the NNS method.

A graphical representation of this discretization procedure is given in Fig.~\ref{fig:diagram} for a quasi-2D system with hexagonal symmetry. In the left panel, we show the $\mathbf{q}$ points involved in the sum in Eq.~\ref{eqn:subsampling1} (a \grid{4} $q$ grid is used in this example). The thicker red line denotes the Brillouin zone edge, and each dot on the left panel represents one point where we calculate both the matrix elements $B$ and the screened Coulomb interaction $W$. In the right panel, we show the special treatment for the Voronoi region associated with the $\mathbf{q}{=}0$ point from Eq.~\ref{eqn:subsampling1}. Each dot in the right panel represents a $\mathbf{q}$ point where we compute the screened Coulomb interaction. In this example, we use $N_s=3$ subsampled points.

When the inverse dielectric matrix is anisotropic, i.e.,  $\varepsilon^{-1}(\mathbf{q}{\rightarrow}0)$ depends on the direction of $\mathbf{q}$, there is an additional complication when employing either the uniform sampling (Eqs.~\ref{eqn:Wavg_MC_3d}) or the NNS scheme (Eqs.~\ref{eqn:subsampling1} and~\ref{eqn:subsampling2}). Still, the angular dependence on the screened Coulomb interaction is typically much less important than the radial dependence, and a simple model can effectively capture most of the anisotropy in $\varepsilon^{-1}(\mathbf{q}{\rightarrow}0)$ without additional computational cost.

For quasi-2D systems and in the long wavelength limit, one can show that the head of the inverse longitudinal dielectric matrix can be expressed as
\begin{align}
\varepsilon_{00}^{-1}(\mathbf{q}{\rightarrow}0) = 1 - q \; \hat{q} \cdot \uuline{\alpha} \cdot \hat{q},
\end{align}
where $\uuline{\alpha}$ is a $2\times2$ Hermitian tensor. The eigenvectors $\{ \mathbf{u}_i \}_i$ of $\uuline{\alpha}$ give the principal axes of polarization for the head of the inverse longitudinal dielectric function, and it can be determined from either symmetry considerations of the crystal or from explicit calculations of  $\varepsilon_{00}^{-1}(\mathbf{q}_0)$ along 3 different directions of $\mathbf{q}_0$.

Our goal is to find an optimal direction $\hat{q}_0$ for the subsampled $\mathbf{q}$ vectors such that, for $|\mathbf{q}'|=|\mathbf{q}_0| \rightarrow 0$,
\begin{align}
\frac{1}{2\pi} \int \mathrm{d}\theta' \; \varepsilon_{00}^{-1}(\mathbf{q}') = \varepsilon_{00}^{-1}(\mathbf{q}_0),
\end{align}
which results in a vector $\mathbf{q}_0$ that is parallel to the average of the eigenvectors of $\uuline{\alpha}$, $\hat{q}_0 = \frac{1}{2}(\hat{u}_1 + \hat{u}_2)$.

For the angular average of $\varepsilon^{-1}(\mathbf{q})$ in  Eq.~$\ref{eqn:Wavg_radial}$ to be accurately represented by a single evaluation of the inverse dielectric matrix along an average direction, one should also choose a Voronoi region that is as isotropic as possible. So, one should keep the ratio $b_i/N_{k_i}$ approximately constant for all extended directions $i$, where each $b_i$ and $N_{k_i}$ is a primitive, reciprocal lattice constant and the number of $\mathbf{k}$ points along each direction, respectively. With this geometrical setup, the NNS scheme can be readily employed on systems with anisotropic screening response in the long wavelength limit as long as the NNS is performed in the direction that averages the screening response. This direction can be obtained by either 3 computationally inexpensive evaluations of the head of the inverse dielectric function along different directions, or by symmetry considerations.



We now turn to applying the NNS method as defined in Eqs.~\ref{eqn:subsampling1} and~\ref{eqn:subsampling2} for some systems of interest. We will first discuss the convergence on semiconducting systems having both isotropic and anisotropic dielectric response -- bilayer \MoSe2{} and monolayer black phosprohous --, and on graphene.

\subsection{\label{sec:subsampling_semicond}NNS method applied to semiconductors}

\begin{table*}
    \caption{\label{tab:mose2}Comparison of the uniform sampling of the Brillouin zone with the non-uniform neck subsampling method for bilayer \MoSe2{} with $N_s=10$ subsampling points. We compare the indirect $\Gamma \rightarrow \Lambda$ gap, CPU usage, and the number of $q$ points in the irreducible portion of the Brillouin zone. For the NNS method, we report the effective $q$ grid spanned by the smaller subsampling point. The two calculations with denser $q$ grids were performed with a smaller cutoff and extrapolated according to the process described on the text.}
    \begin{ruledtabular}
    
    \begin{tabular*}{\textwidth}{@{\extracolsep{\fill}}cccccccc@{}}
	~ & \multicolumn{3}{c}{Uniform sampling} & \multicolumn{4}{c}{Non-uniform neck subsampling (NNS) method} \\
	\cline{2-4} \cline{5-8}
	$q$ grid &
		\begin{tabular}{cc}$\Gamma \rightarrow \Lambda$\\(eV)\end{tabular} &
		\begin{tabular}{cc}CPU usage\\(core--hour)\end{tabular} &
		\begin{tabular}{cc}\# of $q$ points\\in irr. BZ\end{tabular} &
		\begin{tabular}{cc}$\Gamma \rightarrow \Lambda$\\(eV)\end{tabular} &
		\begin{tabular}{cc}CPU usage\\(core--hour)\end{tabular} &
		\begin{tabular}{cc}\# of $q$ points\\in irr. BZ\end{tabular} &
		\begin{tabular}{cc}Effective\\$q$ grid\end{tabular} \\
    \hline
	\grid{6}\phantom{$^\ast$} & 3.31 & 96 & 7 & 1.75 & 157 & 16 & \grid{1143} \\
	\grid{12}\phantom{$^\ast$} & 2.14 & 930 & 19 & 1.76 & 1373 & 28 & \grid{2286} \\
	\grid{24}$^\ast$ & 1.85 & 3620 & 61 & 1.76 & 5130 & 70 & \grid{4573} \\
	\grid{36}$^\ast$ & 1.80 & 12280 & 127 & 1.76 & 15390 & 136 & \grid{6859} \\
    \end{tabular*}
    \end{ruledtabular}
    
    
\end{table*}

\begin{figure}[tp]
   \includegraphics[width=246.0pt]{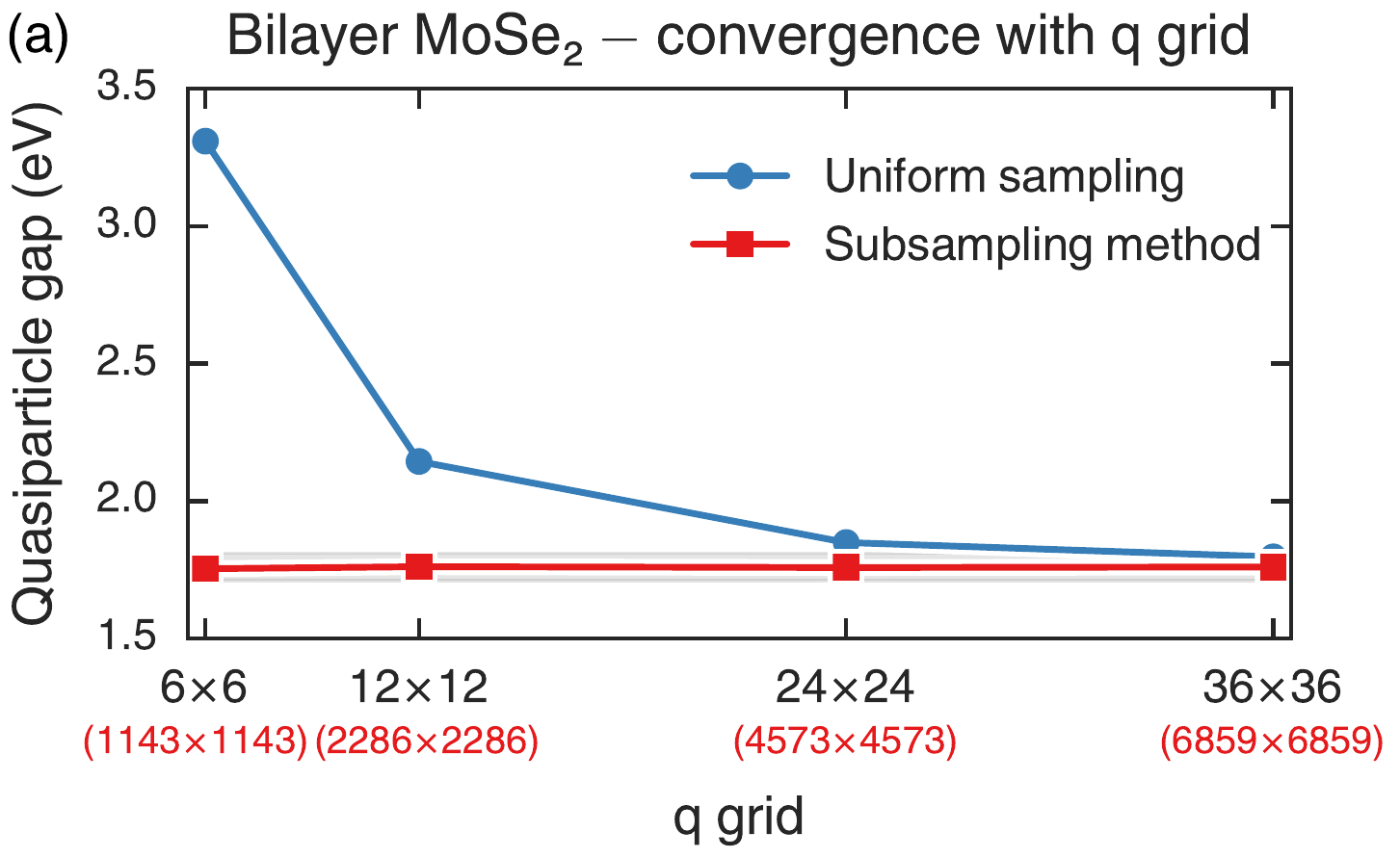}
   \includegraphics[width=246.0pt]{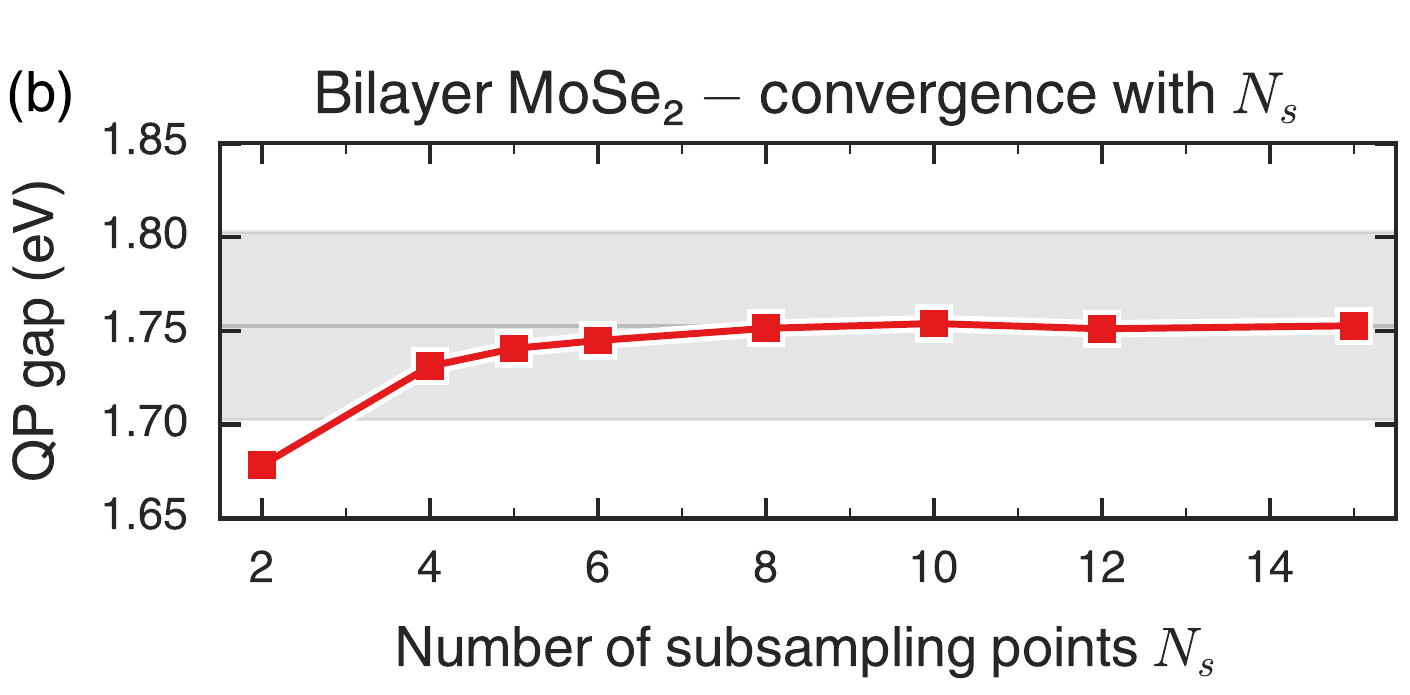}
    \caption{\label{fig:subsampling_MoSe2} (Color online) (a) Convergence of the quasiparticle gap of bilayer \MoSe2 as function of the $q$ grid using a uniform sampling of the Brillouin zone (Monte Carlo averaging scheme) and the proposed NNS method, with $N_s=10$. The values in parenthesis represent the effective $q$ grid captured by the smallest subsampling point. The gray shaded region corresponds to an interval of $\pm50$~meV compared to our most converged value. (b) Converge of the quasiparticle gap as a function of $N_s$ for a $\grid{6}$ $q$ grid. }
\end{figure}

The application of the NNS method is straightforward for systems with isotropic in-plane dielectric response (i.e., such that $\varepsilon^{-1}_\mathbf{G\,G'}(\mathbf{q}{\rightarrow}0)$ does not depend on the direction of $\mathbf{q}$). In Fig.~\ref{fig:subsampling_MoSe2}~(a), we compare the convergence of the quasiparticle gap of bilayer \MoSe2 with the number of $\mathbf{q}$ vectors on a regular grid, using both a conventional sampling of the Brillouin zone with the Monte Carlo sampling method and the new NNS method. We chose bilayer \MoSe2 instead of monolayer \MoSe2 because the gap is indirect in the bilayer structure and converges slightly slower with the number of $\mathbf{q}$ vectors. Whereas one would require a sampling of the $q$-vectors on a \grid{36} grid to converge the quasiparticle gap of bilayer \MoSe2 to within 50~meV using a uniform sampling of the Brillouin zone -- and even finer grids to check that the answer is indeed converged -- we can achieve a much better convergence by sampling the $\mathbf{q}$ vectors on a \grid{6} regular grid with an additional set of $N_s=10$ subsampled $\mathbf{q}$ points, which effectively samples features that would only  be captured on a regular \grid{1143} $q$ grid.

We also compare the convergence of QP gap the as a function of the number of subsampling points in Fig.~\ref{fig:subsampling_MoSe2}~(b), where it is evident that the NNS method converges very fast with the number of subsampling points. Indeed, the quasiparticle gap of bilayer \MoSe2 changes by just $3$~meV if we vary $N_s$ from $8$ to $15$. The extra cost associated with the NNS scheme is also very small in this system, as shown in Table~\ref{tab:mose2}. Therefore, the NNS method allows one to converge the quasiparticle gap of bilayer \MoSe2 in an efficient way, providing savings of about 2 orders of magnitude in the CPU time compared to a traditional uniform sampling  of the Brilluoin zone.

\begin{figure}[tp]
   \includegraphics[width=246.0pt]{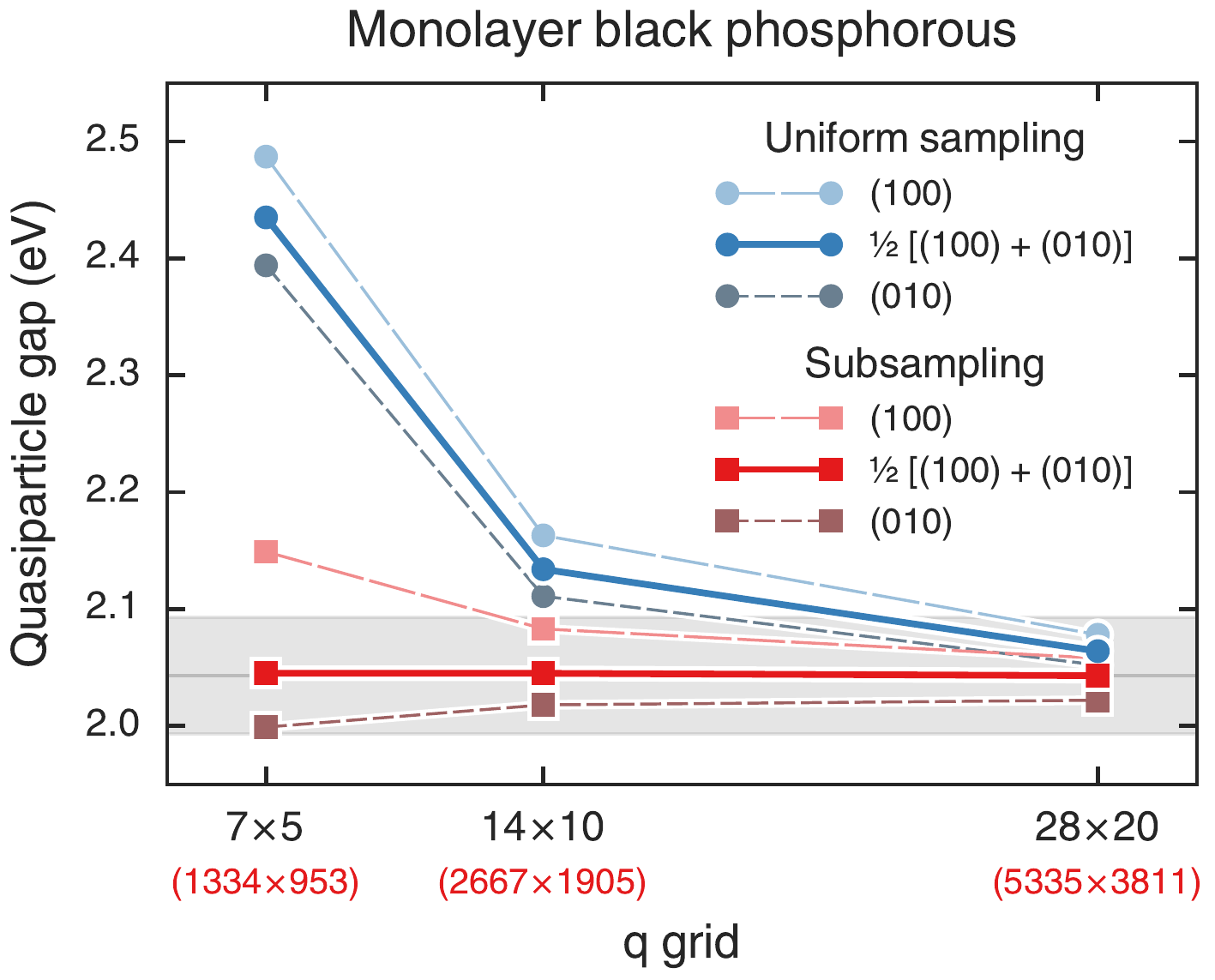}
    \caption{\label{fig:subsampling_P} (Color online) Convergence of the quasiparticle gap of monolayer black phosphorous as function of the $q$ grid using a uniform sampling of the Brillouin zone (Monte Carlo averaging scheme) and the NNS method. For both methods, we compare the converge rate for different directions of the small $\mathbf{q}$ vector(s), where $(100)$ and $(010)$ follow the armchair and zigzag directions, respectively. The gray shaded region corresponds to an interval of $\pm50$~meV compared to our most converged value, and corresponds to the convergence threshold one is typically interested. The values in parenthesis represent the effective $q$ grid captured by the subsampling points.}
\end{figure}

Next, we illustrate the convergence for materials with anisotropic dielectric response by studying monolayer black phosphorous, which is another prototypical quasi-2D semiconductor which exhibits large optical anisotropy, linear optical dichroism, and strong many-body interactions. By the symmetry of the crystal, the principal axes of $1/\varepsilon^{-1}_{00}(\mathbf{q})$, i.e., the eigenvectors of $\uuline{\alpha}$, have to lie along high-symmetry lines. If we setup the lattice such that $(100)$ and $(010)$ correspond to the armchair and zigzag directions of the the black phosphorous monolayer, respectively, we find that the dielectric response is indeed anisotropic in this material, with the eigenvalues of $\uuline{\alpha}$ being different along the two directions: $\alpha_{(100)}\approx 52.6$~1/bohr and $\alpha_{(010)}\approx 72.1$~1/bohr. The optimal direction to compute the dielectric response is $\hat{q}_0=\frac{1}{2}\left[(100) + (010)\right]$, which does not coincide with the $(110)$ direction because the in-plane lattice constants are not the same.

In Fig.~\ref{fig:subsampling_P}, we show the convergence of the GW quasiparticle gap on monolayer black phosphorous as a function of the $q$ grid. Just as in the case of bilayer \MoSe2, we observe that the convergence is much faster with the proposed NNS scheme, where we obtain a quasiparticle gap converged within $50$~meV employing $\mathbf{q}$ vectors on a grid as coarse as ${7}\times{5}\times{1}$ with additional $N_s=10$ subsampled $\mathbf{q}$ points. In addition, we also show that the converge is remarkably fast if we perform the NNS along the optimal direction, $\hat{q}_0 = \frac{1}{2}[(100) + (010)]$. Still, regardless of the direction, the NNS scheme is much more efficient to converge the quasiparticle gap than using a uniform sampling of the Brillouin zone.


\subsection{\label{sec:subsampling_graphene}NNS method applied to quasi-2D metals and quasi-metals}

\begin{figure}[tp]
    \includegraphics[width=246.0pt]{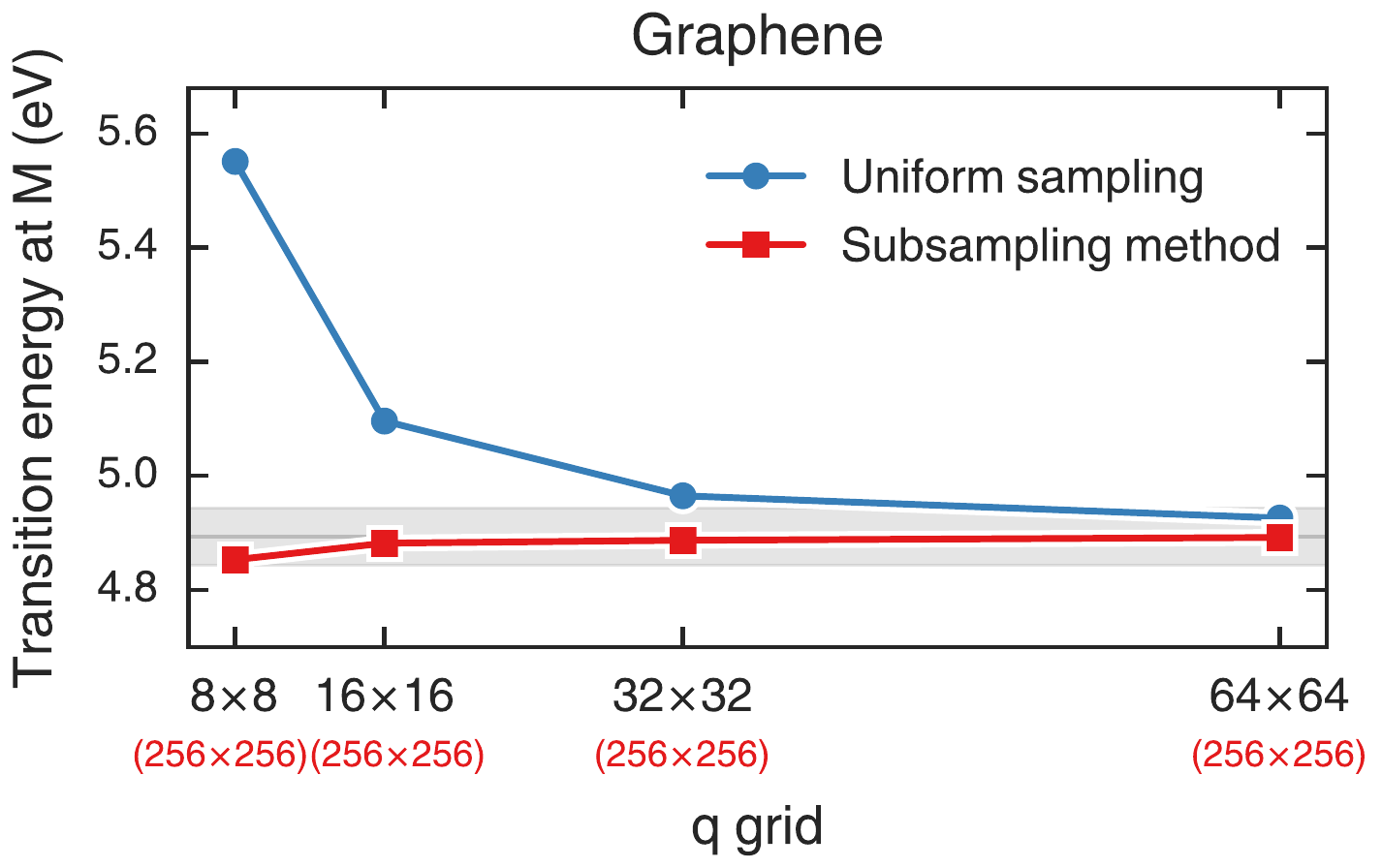}
    \caption{\label{fig:subsampling_graphene} (Color online) Convergence of the quasiparticle interband transition energy at the M point of the Brillouin zone of graphene as function of the $q$ grid. The two curves and the shaded region are the same as in Fig.~\ref{fig:subsampling_MoSe2}. The values in parenthesis represent the effective $q$ grid captured by the subsampling points, which was kept fixed in this calculation.}
\end{figure}

The NNS scheme can also be efficiently applied on systems other than 2D semiconductors. Before we proceed, we must carefully distinguish $\mathbf{k}$ points used to compute the dielectric matrix from the set of transfer momenta $\mathbf{q}$ where we evaluate the dielectric matrix. While the NNS scheme deals with the sampling of $q$-vectors, we have used so far a uniform $k$ grid when we compute $\varepsilon_{\mathbf{G G'}}^{-1}(\mathbf{q})$ for each particular $\mathbf{q}$. However, there is a known additional difficulty when calculating the dielectric matrix for metallic systems because the $k$ grid must be fine enough to sample intraband transitions. This problem can be mitigated by also sampling the $\mathbf{k}$ points in the Brillouin zone in a non-uniform fashion. We write the polarizability matrix as
\begin{widetext}
\begin{align}
\label{eqn:pol2}
\chi_\mathbf{G G'}^0(\mathbf{q},\omega) &=
	\frac{g_s}{V_\mathrm{cell}}
	\sum_{n n' \mathbf{k}} w_\mathbf{k} 
	\left[ f(E_{n' \mathbf{k+q}}) - f(E_{n \mathbf{k}}) \right]
	\frac { M_\mathbf{G}^*(n', n, \mathbf{k, q}) M_\mathbf{G'}(n', n, \mathbf{k, q})}
	{\omega - (E_{n \mathbf{k}} - E_{n' \mathbf{k+q}}) + i0^{+}\,\mathrm{sgn} (E_{n \mathbf{k}} - E_{n' \mathbf{k+q}})}\\
\label{eqn:M}
M_\mathbf{G}(n', n, \mathbf{k, q}) &= \braket{u_{n' \mathbf{k+q}} | e^{i\mathbf{G\cdot r}} | u_{n \mathbf{k}} },
\end{align}
\end{widetext}
where $V_\mathrm{cell}$, $g_s$, $f$, and $w_\mathbf{k}$ denote the unit cell volume, the spin degeneracy of the calculation, the Fermi occupation factor, and the weight of each $\mathbf{k}$ point, with $\sum_\mathbf{k}w_\mathbf{w}=1$, respectively.

Here, we propose to associate different weights, $w_\mathbf{k}$, with each $\mathbf{k}$ point, proportionally to the volume $V_\mathbf{k}$ of the Voronoi cell \vorocell{k} that surrounds each $\mathbf{k}$ point. The weights $w_\mathbf{k}$ can be determined uniquely by the Voronoi tessellation of the Brillouin zone, and we use the Voro++ package in BerkeleyGW~\cite{jdeslip11BGW} to efficiently compute the Voronoi tessellation including periodic boundary conditions. This allows one to use non-uniform $k$ grids to evaluate the sum in Eq.~\ref{eqn:pol2} and more efficiently capture complicated regions of the Brillouin zone, such as those associated with intraband transitions~\cite{voro_pp}.

With this new method, we can now employ the NNS scheme on graphene, another prototypical 2D material where the $k$-point sampling is also complicated. We employ the new non-uniform $k$-point sampling scheme of the dielectric matrix by including $\mathbf{k}$ points from a coarse \grid{8} grid if $\mathbf{k}$ is far away from the Dirac points at the $K$ and $K'$ points of the Brillouin zone but include more $\mathbf{k}$ points commensurate with a much finer $\grid{512}$ grid near the Dirac points. When we lay out the radial subsampling $\mathbf{q}$ points, we employ a constant thicknesses $\Delta_s=\Delta_1$, as we can then reuse more information when we construct the polarizability matrix for different subsampling $\mathbf{q}_s$ vectors. In Fig.~\ref{fig:subsampling_graphene}, we compare the convergence of the quasiparticle interband transition energy at the M point of graphene with the number of $q$-vectors on the regular grid, for the two methods. Once again, the curve obtained with the proposed NNS method converges much faster with the number of $q$-vectors, and a regular \grid{8} grid for the $q$-vectors is enough to converge the quasiparticle transition energy at the M point to within 50~meV.


Therefore, the NNS method can indeed be applied on a variety of systems with reduced dimensionality and with different screening properties. Although the method was tested here for GW calculations, which show order-of-magnitude speedups in the computer runtime compared to a regular sampling of the Brillouin zone, it can be applied to other kinds of many-electron perturbation theory calculations as well.



\section{\label{sec:clustered}Clustered sampling interpolation (CSI) method}

In the previous section, we introduced the NNS method to compute sums associated with one-electron integrals, such as those needed to calculate the GW quasiparticle self energy. In this section, we introduce the clustered sampling interpolation (CSI) method to efficiently sample the Brillouin zone for problems involving two-particle correlated states, such as those given by the Bethe-Salpeter equation (BSE). These problems are characterized by associated Hamiltonians, typically written in the occupation representation, with matrix elements describing scattering amplitudes from one two-particle state at a $\mathbf{k}$ point to another state at a different $\mathbf{k}$ point.

When solving the BSE to obtain the optical absorption spectrum, it is a well-known problem that interaction matrix elements need to be constructed on a very fine $k$ grid because excitons are correlated states with wavefunction that has fine structures in k-space. For example, even for bulk semiconductors such as GaAs, very fine uniform grids containing over a million $\mathbf{k}$ points are necessary to resolve the exciton energies and wavefunctions~\cite{rohlfing98}. In many cases, the bottleneck for solving the BSE is in computing the interaction matrix elements, and in the past, schemes based on interpolation between two different $k$ grids, which we refer to as "dual-grid" methods, have been employed to make these calculations feasible~\cite{rohlfing98,rohlfing00}. We will review here the scheme implemented in the current released version of BerkeleyGW~\cite{jdeslip11BGW}, describe its shortcomings when applying it on systems with reduced dimensionality, and propose an extension for the scheme to mitigate these shortcomings.

We are interested in evaluating the two-particle matrix elements that are in the Bethe-Salpeter equation, which is of the form
\begin{equation}
\begin{split}
\label{eqn:BSE}
&(E_{c\textbf{k}+\textbf{Q}}-E_{v\textbf{k}})A^{S}_{vc\textbf{k}} + \\ &\sum_{v'c'\textbf{k}'}
K^\mathrm{eh}_{vc;v'c'}(\mathbf{k}, \mathbf{q}=\mathbf{k}'{-}\mathbf{k})
A^S_{v'c'\textbf{k}'} = \Omega^S A^{S}_{vc\textbf{k}}.
\end{split}
\end{equation}
Here, $S$ indexes the exciton states; $\mathbf{Q}$ is the center-of-mass momentum of the electron-hole pair; $A^{S}_{vc\textbf{k}}$ is the amplitude of a free electron-hole pair consisting of an electron in $|c\textbf{k}+\textbf{Q}\rangle$ and one missing from $|v\textbf{k}\rangle$; $\Omega^S$ is the exciton excitation energy; $E_{c\textbf{k}+\textbf{Q}}$ and  $E_{v\textbf{k}}$ are the quasiparticle energies, and $K^{\mathrm{eh}}$ is the electron-hole interaction kernel. The kernel contains contributions from a direct term and an exchange term. The direct term is
\begin{equation}
\label{eqn:Kd}
\begin{split}
K^\mathrm{d}_{vc;v'c'}&(\mathbf{k}, \mathbf{q}=\mathbf{k}'{-}\mathbf{k}) =\\
	\sum_{\mathbf{G G'}} 
	M_{\mathbf{G}}^{*}(c, c', \mathbf{k, q}) &
	W_{\mathbf{G, G'}}(\mathbf{q}) \,
	M_{\mathbf{G}'}(v, v', \mathbf{k, q}),
\end{split}
\end{equation}
where the matrix elements $M$ are given by Eq.~\ref{eqn:M}, and the exchange term is
\begin{equation}
\label{eqn:Kx}
\begin{split}
K^\mathrm{x}_{vc;v'c'}&(\mathbf{k}, \mathbf{q}=\mathbf{k}'{-}\mathbf{k}) = \\
	\sum_{\mathbf{G}} 
	M_{\mathbf{G}}^{*}(c, v, \mathbf{k, Q}) &
	v_{\mathbf{G}}(\mathbf{Q}) \,
	M_{\mathbf{G}}(c', v', \mathbf{k}',\mathbf{Q}).
\end{split}
\end{equation}

In the BerkeleyGW code package, the original formulation of the dual-grid interpolation method employs two sets of $\mathbf{k}$ points: one set of $\mathbf{k}_\mathrm{co}$ $\mathbf{k}$ points defined on a coarse grid and a set of $\mathbf{k}_\mathrm{fi}$ $\mathbf{k}$ points defined on a fine grid. The direct ($K^{\mathrm{d}}$) and exchange ($K^{\mathrm{x}}$) matrix elements in the BSE kernel are explicitly calculated on $\mathbf{k}_\mathrm{co}$. Then, an interpolation is performed by expanding each fine-grid Bloch state in terms of the closest coarse-grid Bloch state,
\begin{align}
\label{eqn:bloch}
\ket{u_{n\mathbf{k}_{\mathrm{fi}}}} &= \sum_{m}
	C_{nm}^{\mathbf{k}_{\mathrm{co}}} \ket{u_{m\mathbf{k}_{\mathrm{co}}}} \\
C_{nm}^{\mathbf{k}_{\mathrm{co}}} &= \int \mathrm{d}^3r \;
	u_{n\mathbf{k}_{\mathrm{fi}}}(\mathbf{r}) \, u_{m\mathbf{k}_{\mathrm{co}}}^{*}(\mathbf{r}),
\end{align}
which allows the kernel matrix elements to be approximated as
\begin{equation}
\begin{split}
&K^\mathrm{d/x}_{mn;m'n'}(\mathbf{k}_\mathrm{fi},
	\mathbf{q}_\mathrm{fi}=\mathbf{k}_\mathrm{fi}'{-}\mathbf{k}_\mathrm{fi})
\approx \\
	\sum_{\substack{n_1 n_2 \\ m_1 m_2}} \,
	C_{n n_1}^{\mathbf{k}_\mathrm{co}} \,&
	C_{m m_1}^{\mathbf{k}_\mathrm{co}*} 
	C_{n'n_2}^{\mathbf{k}_\mathrm{co}'*} \,
	C_{m'm_2}^{\mathbf{k}_\mathrm{co}'} \;
K^\mathrm{d/x}_{mn;m'n'}(\mathbf{k}_\mathrm{fi},
	\mathbf{q}_\mathrm{co}=\mathbf{k}_\mathrm{co}'{-}\mathbf{k}_\mathrm{co}).
\end{split}
\end{equation}

Notice in Eq.~\ref{eqn:Kd} and Eq.~\ref{eqn:Kx} that $K^{\mathrm{d}}$ depends sensitively on the relative reciprocal vector $\mathbf{q}$, whereas $K^{\mathrm{x}}$ depends only on the center-of-mass momentum $\mathbf{Q}$, which is a constant. The bare Coulomb interaction $v(\mathbf{q})$ diverges as $\mathbf{q}\rightarrow 0$.
We therefore expect that $K^{\mathrm{d}}$ will change very rapidly at small $\mathbf{q}$, and consequently, any direct interpolation of $K^{\mathrm{d}}$ must converge very slowly. To avoid this problem, as currently implemented in BerkeleyGW, $K^{\mathrm{d}}$ is decomposed into its head ($K^\mathrm{h}$), wing ($K^\mathrm{w}$) and body ($K^\mathrm{b}$) contributions, each of which has different limiting behavior for the Coulomb interaction as $\mathbf{q}\rightarrow 0$. The head contains the terms where $\mathbf{G}{=}\mathbf{G}'{=}0$ and diverges as $\frac{1}{q^2}$ in 3D and $\frac{1}{q}$ in 2D, for semiconductors and insulators. The wing contains the sum over terms where $\mathbf{G}{=}0\neq\mathbf{G}'$ or $\mathbf{G}'{=}0\neq\mathbf{G}$ and diverges as $\frac{1}{q}$ in 3D and goes to a constant in 2D, for semiconductors and insulators. The body contains the sum over terms where $\mathbf{G}\neq 0$ and $\mathbf{G}'\neq 0$ and goes to a constant value in the limit of small $q$. With this understanding, for the 2D and 3D cases, the direct term can be written as
\begin{equation}
\begin{split}
&K^\mathrm{d}_{mn;m'n'}(\mathbf{k}, \mathbf{q}) = \\
&\frac{a_{mn;m'n'}(\mathbf{k}, \mathbf{q})}{q^{d-1}} +
\frac{b_{mn;m'n'}(\mathbf{k}, \mathbf{q})}{q^{d-2}} +
c_{mn;m'n'}(\mathbf{k}, \mathbf{q}),
\end{split}
\end{equation}
where $d$ is the effective dimension, and 
\begin{equation}
\begin{split}
a_{mn;m'n'}(\mathbf{k}, \mathbf{q})&= q^{d-1}\times K^\mathrm{h}_{mn;m'n'}(\mathbf{k}, \mathbf{q}),
\\
b_{mn;m'n'}(\mathbf{k}, \mathbf{q})&= q^{d-2}\times K^\mathrm{w}_{mn;m'n'}(\mathbf{k}, \mathbf{q}),
\\
c_{mn;m'n'}(\mathbf{k}, \mathbf{q})&= K^\mathrm{b}_{mn;m'n'}(\mathbf{k}, \mathbf{q}).
\end{split}
\end{equation}
The functions $a$, $b$, and $c$, where the divergence in the Coulomb interaction is removed, are interpolated in the dual-grid scheme and then used to construct $K^{d}$.

\begin{figure}[tp]
    \includegraphics[width=246.0pt]{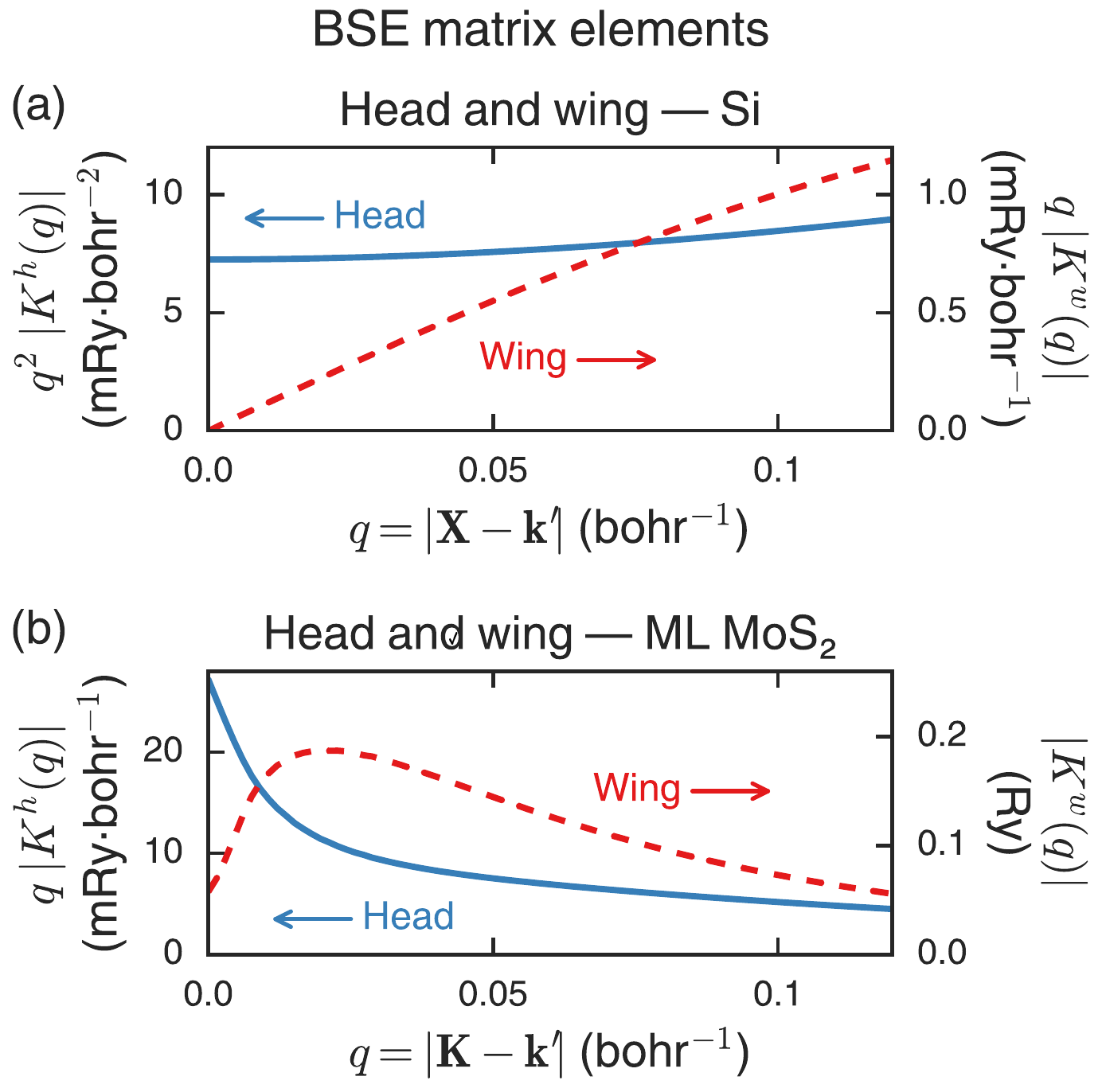}
    \caption{\label{fig:mtxel} (Color online) Head and wing components of the BSE matrix elements for bulk Si (top) and monolayer \MoS2 (bottom).}
\end{figure}

The interpolation procedure described above works efficiently for 3D metals and semiconductors, where $a$, $b$, and $c$ are smooth functions of $\mathbf{q}$ because the inverse dielectric matrix is also a smooth function of $\mathbf{q}$. However, as we previously discussed, $\varepsilon^{-1}_{\mathbf{G G'}}(\mathbf{q})$ displays sharp features in systems with reduced dimensionality as $\Qo$, and thus, the head, wing and body components of the matrix elements, which depend on $\varepsilon^{-1}$, may also vary considerably with $\mathbf{q}$ even when the divergence in the Coulomb interaction is removed. We illustrate this behavior by plotting the head and wing components of the matrix elements associated with the direct Coulomb term of the BSE (Eq.~\ref{eqn:Kd}) for silicon and monolayer \MoS2 on Fig.~\ref{fig:mtxel}. In bulk silicon, we multiply the head matrix elements by $q^2$ and the wing matrix elements by $\mathbf{q}$ to remove the divergence due to the bare Coulomb interaction. Then, the matrix elements are smooth functions of $\mathbf{q}$. In 2D, however, the non-smooth behavior cannot be removed by multiplying any simple factor. We remove the divergence due to the bare Coulomb interaction as by multiplying the head matrix element by $q$, but even after removing the Coulomb divergence, both the head and wing components of still have a sharp features at small $\mathbf{q}$. These features are a consequence of the sharp feature in the inverse dielectric matrix (Fig.~\ref{fig:epsinv}).

To capture these sharp features in 2D, it is important to explicitly calculate $K^\mathrm{h/w/b}_{mn;m'n'}(\mathbf{k}, \mathbf{q})$ for a variety of small $\mathbf{q}$. Consequently, a dual-grid scheme as described above necessarily converges very slowly with respect to sampling of the coarse grid, which must be fine enough to resolve the sharp feature in $K^{\mathrm{h/w/b}}$, and quickly becomes prohibitively expensive, since the cost of calculating the matrix elements scales with the number of coarse $\mathbf{k}$ points squared.

\begin{figure}[tp]
    \includegraphics[width=246.0pt]{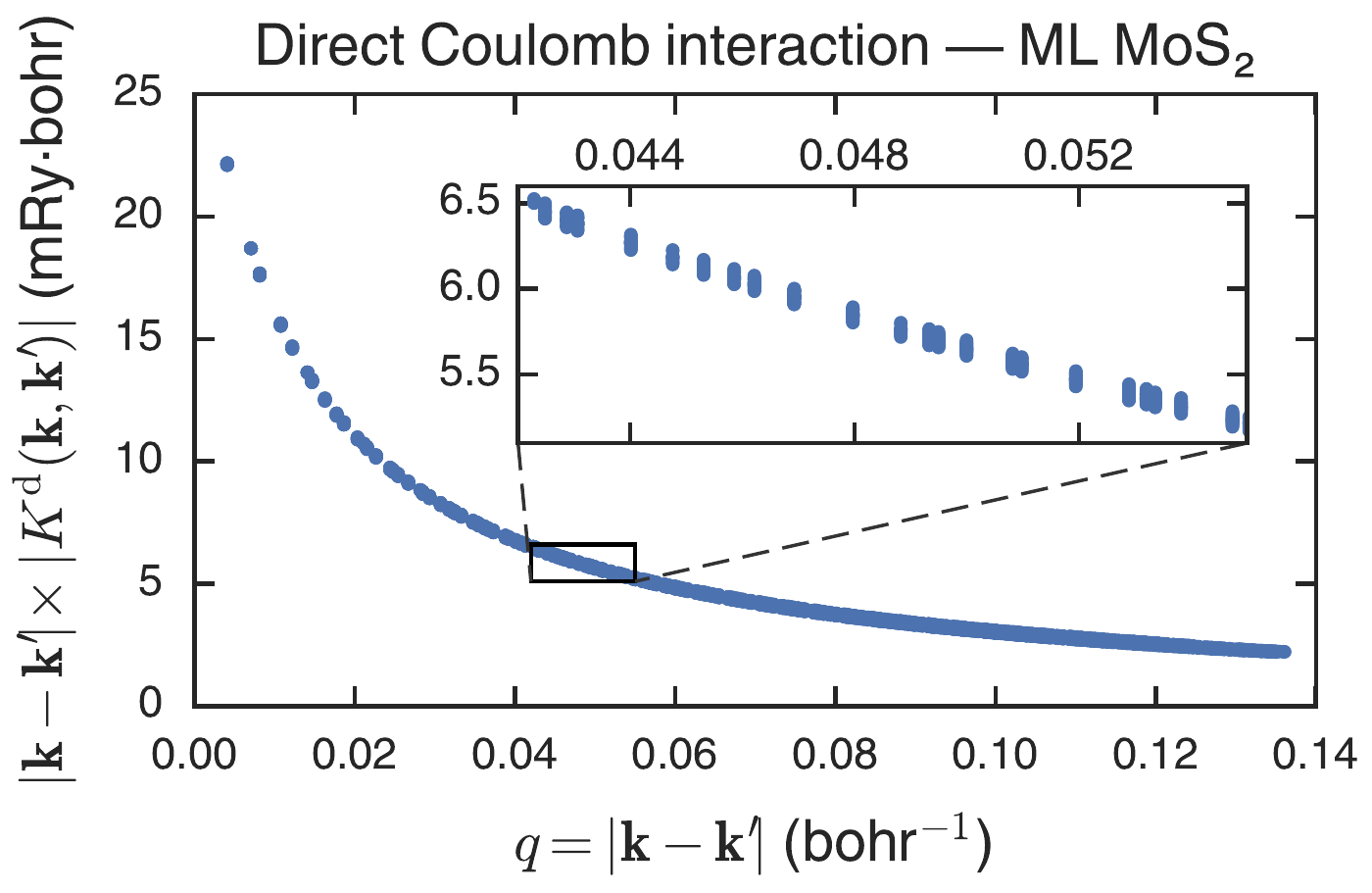}
    \caption{\label{fig:Wiso} (Color online) Matrix element for the BSE involving the direct screened Coulomb interaction, $K^\mathrm{d}(\mathbf{k},\mathbf{k}')$, calculated for an electron and hole states at the conduction band and valence band, respectively, for monolayer \MoS2, where $\mathbf{k}$ and $\mathbf{k}'$ are any points on a \grid{300} $k$ grid that lies within 0.02 bohr$^{-1}$ of the K point. Note that $K^\mathrm{d}(\mathbf{k},\mathbf{k}')$ depends very weakly on either $\mathbf{k}$ or $\mathbf{k}'$ individually if $\mathbf{q}=\mathbf{k}'-\mathbf{k}$ is kept constant. The inset zooms in on a small region of the plot and illustrates the small spread in values for different $\mathbf{k}$ and $\mathbf{k}'$ pairs with the same $q$.}
\end{figure}

In contrast to their sharply varying $q$-dependence, however, the head, wing, and body matrix elements do not depend much on $\mathbf{k}$. This is because the $k$-dependence comes in solely in the matrix elements $M(m,n,\mathbf{k,q})$, which are typically smooth functions of $\mathbf{k}$, since the periodic part of the Bloch functions are smoothly varying quantities. This is illustrated for the case of monolayer \MoS2 in Fig.~\ref{fig:Wiso}, where the contribution to the BSE from the direct screened Coulomb interaction, $K^\mathrm{d}(\mathbf{k},\mathbf{k}'{=}\mathbf{k{+}q})$, displays a very small spread over a wide range of values for $\mathbf{k}$. Thus, in order to capture all of the screening effects, we need to minimally sample a large number of finely-spaced $\mathbf{q}$ transitions from a set of $\mathbf{k}$ points that can be relatively coarse.

In order to explicitly capture the small-$\mathbf{q}$ behavior, we develop an extension on the dual-grid interpolation scheme. In addition to calculating BSE matrix elements on a coarse $k$ grid, we also explicitly calculate BSE matrix elements for scattering from each $\mathbf{k}$ point, $\mathbf{k}_\mathrm{co}$, on the coarse grid to an arbitrary cluster of $\mathbf{k}$ points, $\mathbf{k}_\mathrm{cl}$, close to each $\mathbf{k}_\mathrm{co}$. We will refer to this scheme as clustered sampling interpolation (CSI). For simplicity, we will focus on the case of isotropic materials, where $K^\mathrm{h/w/b}_{mn;m'n'}(\mathbf{k}, \mathbf{q})$ depends only on $|\mathbf{q}|$. In this case, the cluster of points can be chosen to lie along a radial line extending from each $\mathbf{k}_\mathrm{co}$. The generalization to anisotropic materials is straightforward with a small computational overhead.

We interpolate the BSE matrix elements from the coarse grid to the fine grid using a conditional scheme. If the distance between two points on the fine grid, $|\mathbf{k}'_{\mathrm{fi}}-\mathbf{k}_{\mathrm{fi}}|$, is greater than the smallest distance between two points on the coarse grid $\Delta_\mathrm{co}$, the interpolation is identical to the original dual-grid scheme. If the distance between two points on the fine grid, $|\mathbf{k}'_{\mathrm{fi}}-\mathbf{k}_{\mathrm{fi}}|$, is less than $\Delta_\mathrm{co}$, we expand the Bloch state at $\mathbf{k}_{\mathrm{fi}}$ over the Bloch states at the closest coarse point, $\mathbf{k}_{\mathrm{co}}$, and we expand the Bloch state at $\mathbf{k}_{\mathrm{fi}}'$ over the Bloch states at a cluster point, $\mathbf{k}_{\mathrm{cl}}$, for which $K^\mathrm{h/w/b}_{mn;m'n'}(\mathbf{k}_\mathrm{co}, \mathbf{q}=\mathbf{k}_\mathrm{cl}-\mathbf{k}_\mathrm{co})$ has already been calculated, while preserving as closely as possible the length of the transfer vector so that $|\mathbf{q}|=|\mathbf{k}_{\mathrm{co}}-\mathbf{k}_{\mathrm{cl}}|\approx|\mathbf{k}'_{\mathrm{fi}}-\mathbf{k}_{\mathrm{fi}}|$. Then,
\begin{equation}
\begin{split}
\label{eqn:csi}
K^\mathrm{h/w/b}_{mn;m'n'}&(\mathbf{k}_\mathrm{fi},
	\mathbf{q}_\mathrm{fi}{=}\mathbf{k}'_\mathrm{fi}{-}\mathbf{k}_\mathrm{fi})
\approx \\
	\sum_{\substack{n_1 n_2 \\ m_1 m_2}} 
	C_{n n_1}^{\mathbf{k}_\mathrm{co}} 
	C_{m m_1}^{\mathbf{k}_\mathrm{co}*} 
	C_{n'n_2}^{\mathbf{k}_\mathrm{cl}*} &
	C_{m'm_2}^{\mathbf{k}_\mathrm{cl}} \; 
K^\mathrm{h/w/b}_{mn;m'n'}(\mathbf{k}_\mathrm{fi},
	\mathbf{q}_\mathrm{co}={\mathbf{k}_\mathrm{cl}}{-}\mathbf{k}_\mathrm{co}),
\end{split}
\end{equation}
where
\begin{equation}
C_{nm}^{\mathbf{k}'_{\mathrm{cl}}} = \int \mathrm{d}^3r \;
	u_{n\mathbf{k}'_{\mathrm{fi}}}(\mathbf{r}) \, u_{m\mathbf{k}_{\mathrm{cl}}}^{*}(\mathbf{r}).
\end{equation}

\begin{figure}[htbp]
    \includegraphics[width=246.0pt]{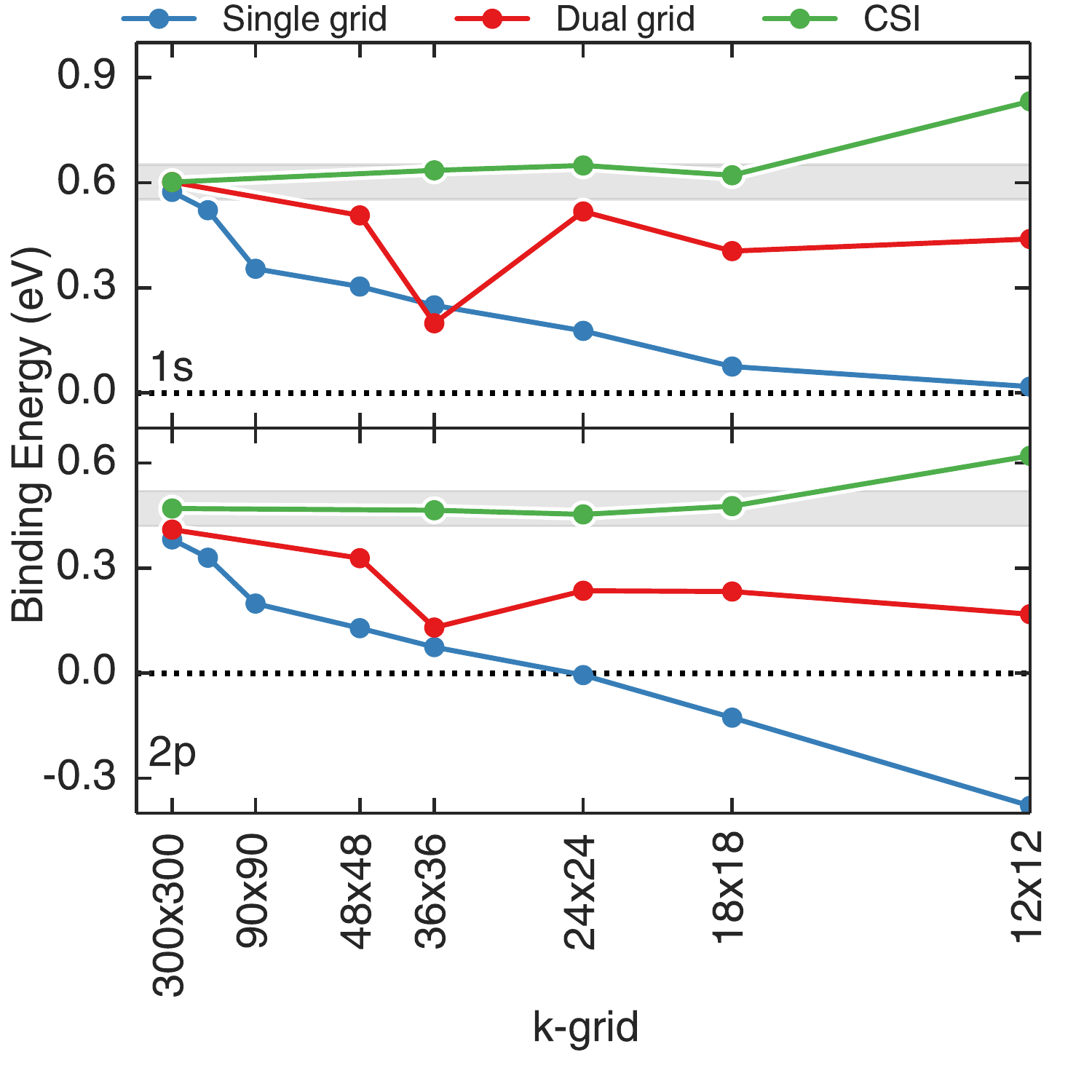}
    \caption{\label{fig:csiconverge} (Color online) Convergence of the binding energy of the 1s(top) and 2p(bottom) states of the lowest energy series of excitons in \MoS2 using an explicit calculation without interpolation (single grid), a dual-grid method and the proposed CSI method. The x-axis represents the $k$-point grid used in the single grid method and the coarse grid used in both the dual grid and CSI methods. The gray shaded region corresponds to an interval of $\pm 50$ meV compared to the converged value, which is the convergence threshold one is typically interested in.}
\end{figure}

We now apply clustered sampling interpolation to a system of interest. Fig.~\ref{fig:csiconverge} shows how the binding energy of the lowest energy 1s and 2p excitons in monolayer \MoS2 converges with respect to an explicit calculation on a single uniform grid (the single-grid scheme) and with respect to the coarse $k$ grid when using either dual-grid interpolation or clustered sampling interpolation. For both the dual-grid interpolation and CSI, the coarse grid is interpolated to a \grid{300} fine grid. The binding energy is defined, following Ref.~\cite{qiu16}, as the difference between the electron-hole continuum and the exciton excitation energy, which is independent of the numerical treatment of the divergence at $W(\textbf{q}=0)$. From Fig.~\ref{fig:csiconverge}, it is clear that the clustered sampling interpolation converges much more quickly than the dual-grid interpolation, requiring only a \grid{18} coarse grid to converge the binding energy to within 0.1~eV. In contrast, the dual-grid scheme does not converge until the coarse grid sampling is increased beyond \grid{48}. Moreover, we see that while in most cases the dual-grid interpolation is still an improvement on the uniform gird, the convergence fluctuates. This erratic convergence occurs because different uniform $k$ grids sample different regions of the sharp feature in the screening. The different convergence rates are even more dramatic for higher energy states, such as the 2p state (Fig.~\ref{fig:csiconverge}), whose complex nodal structure is even more sensitive to the spatially varying screening at small $q$.

In general, calculating the BSE matrix elements scales with the total number of $\mathbf{k}$ points squared. Thus, in the dual-grid scheme, the computational cost scales with the number of $\mathbf{k}$ points on the coarse grid squared, $N_{\mathbf{k}_{\mathrm{co}}}^2$. Clustered sampling interpolation has an additional cost associated with calculating the matrix elements involving transitions between the coarse $\mathbf{k}$ points and cluster points. This additional term scales as $N_{\mathbf{k}_{\mathrm{co}}}\times N_{\mathbf{k}_\mathrm{cl}}$, where $N_{\mathbf{k}_\mathrm{cl}}$ is the number of $\mathbf{k}$ points in each cluster. For isotropic systems $N_{\mathbf{k}_\mathrm{cl}}$ is typically much smaller than $N_{\mathbf{k}_{\mathrm{co}}}$, since the sampling is only along one dimension. Thus, the additional cost of clustered sampling interpolation is small compared with the cost of calculating the matrix elements on the coarse grid. The total cost of calculating the matrix elements scales as $N_{\mathbf{k}_{\mathrm{co}}}^2+N_{\mathbf{k}_{\mathrm{co}}}\times N_{\mathbf{k}_\mathrm{cl}}$.

Table~\ref{tab:csi} shows the CPU time required to calculate the BSE kernel,  $K^{\mathrm{eh}}$, for MoS$_{2}$ in the dual-grid and CSI schemes as a function of the coarse grid and interpolated to a \grid{300} fine grid. On identical coarse grids, there is a small computational overhead in the CSI scheme, on the order of 10 core-hours, which scales linearly with $N_{\mathbf{k}_{\mathrm{co}}}$. However, the binding energy of the 1s exciton state converges in the CSI scheme on an \grid{18} coarse grid, whereas it is still unconverged on a \grid{48} coarse grid in the dual-grid scheme. Thus, for MoS$_{2}$, CSI results in a speed-up of \textit{at least} one order of magnitude. To directly calculate the BSE matrix elements on a \grid{300} as reported in Fig.~\ref{fig:csiconverge}, we solve the BSE on a patch, which only includes $\mathbf{k}$ points within 0.2$\mathrm{\AA}^{-1}$ of the K point in the Brillouin zone. This allows us to obtain the binding energy of the lowest energy excitons within 20meV of the calculation on the full Brillouin zone but is insufficient to obtain the entire optical spectrum. Since we know that calculating the BSE matrix elements scales as $N_{\mathbf{k}_{\mathrm{co}}}^2$, we estimate that directly calculating the BSE matrix on a \grid{300} $k$ grid would take approximately 15 million core-hours, compared with 228 core hours with CSI scheme.

\begin{figure}[htbp]
    \includegraphics[width=246.0pt]{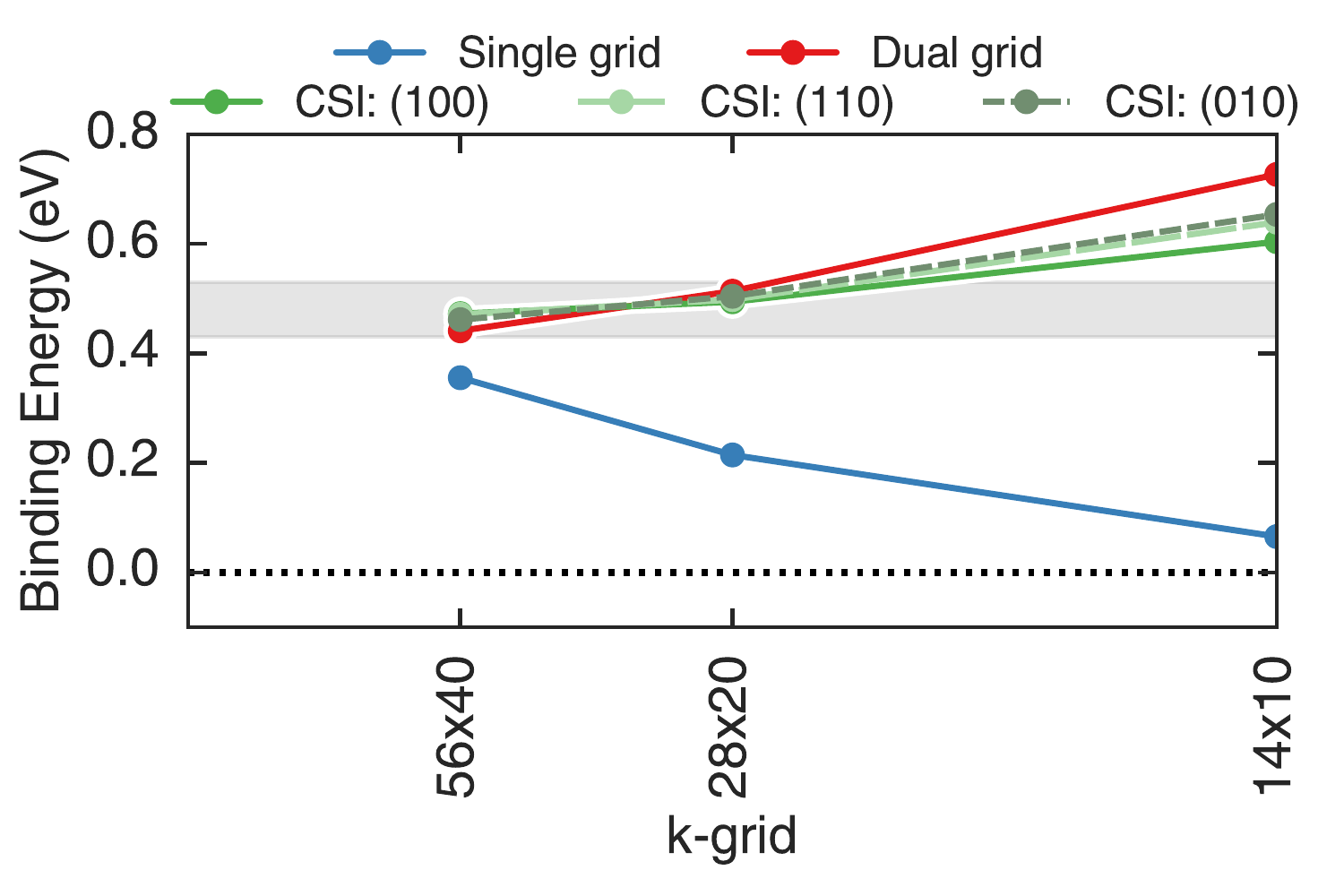}
    \caption{\label{fig:csiP} (Color online) Convergence of the binding energy of the 1s(top) state of the lowest energy series of excitons in monolayer black phosphorus using an explicit calculation without interpolation (single grid), a dual-grid method and the proposed CSI method with sampling along the 100, 110, and 010 directions. The x-axis represents the $k$-point grid used in the single grid method and the coarse grid used in both the dual grid and CSI methods. The gray shaded region corresponds to an interval of $\pm 50$ meV compared to the converged value, which is the convergence threshold one is typically interested in.}
\end{figure}

The proposed clustered sampling scheme assumes an isotropic system, where the BSE matrix elements depend only on the magnitude of $q$. However, in practice, it also results in improved convergence for anisotropic materials such as few-layer black phosphorus. Fig.~\ref{fig:csiP} shows the performance of the CSI scheme for monolayer black phosphorus, when the clustered points are sampled along the $(100)$ direction, $(110)$ direction and $(010)$ direction. Here, the convergence of the binding energy for the different interpolation schemes is referenced to a single grid calculation with $160\times160\times1$ k-points performed in a patch of radius \angs{0.2} around the $\Gamma$ point in the Brillouin zone. The reference converged binding energy is 0.47 eV. For the same coarse and fine $k$ grids, the CSI scheme always converges more quickly than both the single and dual grid scheme, with convergence being the fastest when the clustered points are sampled along the $(100)$ direction, which is the more highly-dispersive armchair direction in black phosphorus. Once again, the convergence of the dual grid scheme is erratic due to the spatially varying screening. While some $k$ grids give similar results to the CSI scheme, as the $k$ grid is increased to $56\times 40\times 1$ the dual grid binding energy still undershoots the converged value.

\begin{table}
    \caption{\label{tab:csi} Comparison of computational time
      required to calculate the BSE kernel matrix elements
      $K^{\mathrm{eh}}$ with different Brillouin Zone
      interpolation schemes for
      MoS$_{2}$ on different coarse $k$ grids, $k_{\mathrm{co}}$. The
      binding energy of the 1s state, $E_{b}^{\mathrm{1s}}$, is given after interpolating from the coarse $k$
      grid to a \grid{300} fine grid.
}
    \begin{ruledtabular}
    
    \begin{tabular*}{\textwidth}{ccccc}
	~ &  \multicolumn{2}{c}{Dual Grid} & \multicolumn{2}{c}{CSI} \\
	\cline{2-3} \cline{4-5} \\
	$k_{\mathrm{co}}$ grid & 
		\begin{tabular}{cc}CPU Usage\\(core--hour)\end{tabular} &
		\begin{tabular}{cc}E$_{b}^{\mathrm{1s}}$\\(eV)\end{tabular} &
		\begin{tabular}{cc}CPU Usage\\(core--hour)\end{tabular} &
		\begin{tabular}{cc}E$_{b}^{\mathrm{1s}}$\\(eV)\end{tabular} \\
    \hline
	\grid{12} &  39 & 0.44 & 59 & 0.83 \\
	\grid{18} &  196 & 0.41 & 223 & 0.62 \\
	\grid{24} &  579 & 0.52  & 613 &  0.65 \\
	\grid{36} &  2948 & 0.20 & 3017 & 0.64 \\
	\grid{48} &  8857 & 0.51  & 8979 & 0.63  \\
    \end{tabular*}
    \end{ruledtabular}
    
\end{table}

\section{Conclusion\label{sec:conclusion}}

In summary, we address the problem that many-electron perturbation theory calculations,  such as those performed in GW and GW-BSE theories, on low-dimensional systems converge very slowly with respect to sampling of the Brillouin zone due to sharp features in the spatial variations in screening, which manifest as sharp features in the $q$-dependence of the dielectric matrix and cannot be described by a simple analytic model due to the complexity of the out-of-plane local fields. Thus, we present two new schemes to sample the Brillouin zone in a computationally efficient way for low-dimensional systems. The first scheme, which we refer to as the non-uniform neck subsampling (NNS) method, allows for efficient sampling of single-particle problems, such as GW and ACFDT. In the NNS method, an additional radial sampling is performed in the Voronoi cell that surrounds each $\mathbf{k}$ point with appropriately chosen weights. The second scheme, clustered sampling interpolation (CSI), addresses two-particle scattering problems, such as in solving for the solution of the BSE. In CSI, we explicitly calculate two-particle scattering matrix elements in small, uniformly-spaced clusters of $\mathbf{k}$ points and use these clusters to interpolate to a uniform fine grid. Both schemes result in typical speedups of about two orders of magnitude in the computer run-time and can be easily incorporated into several \textit{ab initio} packages that compute electronic and optical properties employing many-body-perturbation theory methods.

This work was supported by the Center for Computational Study of Excited State Phenomena in Energy Materials at the Lawrence Berkeley National Laboratory, which is funded by the U. S. Department of Energy, Office of Basic Energy Sciences under Contract No. DE-AC02-05CH11231, and by the by National Science Foundation Grant No. DMR-1508412. D. Y. Q. acknowledges support from the NSF Graduate Research Fellowship Grant No. DGE 1106400. Computational resources have been provided from the Extreme Science and Engineering Discovery Environment (XSEDE), which is supported by National Science Foundation grant number ACI-1053575, and from the National Energy Research Scientific Computing Center (NERSC), a DOE Office of Science User Facility supported by the Office of Science of the U.S. Department of Energy under Contract No. DE-AC02-05CH11231.

\section{\label{sec:Appendix}Appendix}

For a given $N_s$ number of subsampled points, we have the freedom to define two quantities: the $N_s$ annulus thicknesses $\Delta_s$ that define the intervals for the radial integrals and, thus, the weights $w_s$; and the $N_s$ subsampling points $q_s$ where the dielectric matrix has to be explicitly computed. In order to make the discretization in Eq.~\ref{eqn:Wavg_radial} practical, it is necessary to choose appropriate subsampled points $q_s$ and/or the thicknesses $\Delta_s$ that approximates the integral in an efficient way. We introduce a scheme to use a crude approximation of the screened Coulomb interaction to provide constraints on either $q_s$ or $\Delta_s$.

For semiconductors, the inverse dielectric matrix approaches a finite constant as \Qo{}, so we can write the head of the (truncated) screened Coulomb interaction as $W(q) \propto 1/q^2$, $1/q$ or $\log(q)$, depending on whether we have a 3D, 2D or 1D system, respectively. A good choice of $\Delta_s$ and $q_s$ is such that the screened Coulomb interaction evaluated at the subsampled point represents the average value of the $W$ according to the analytic limit,
\begin{align}
\label{eqn:Wavg_opt}
W(\mathbf{q}_s, \omega{=}0) \int_{a_{s}}^{a_{s+1}} \mathrm{d}^D q' =
\int_{a_{s}}^{a_{s+1}} \mathrm{d}^D q' \; W(\mathbf{q}', \omega{=}0),
\end{align}
which, together with the constraint of $a_{N_s}$ from \vorocell{q{=}0}, provide $N_s+1$ constraints and allows us to obtain the optimal subsampling points $q_s$ given $N_s-1$ thicknesses $\Delta_s$ for the radial integration.

We summarize the relationship between $\Delta_s$ and $q_s$ obtained from Eq.~\ref{eqn:Wavg_opt} for 3D, 2D and 1D semiconducting systems in Table~\ref{tab:coefs}. For systems other than semiconductors, the screened Coulomb interaction has different analytic behaviors for \Qo{}, so other optimal subsampling points could be determined. Fortunately, for metallic systems of any dimensionality, the condition in Eq.~\ref{eqn:Wavg_opt} is fulfilled for any choice of $q_s$, and for quasi-2D systems with linear energy dispersion, such as graphene, the optimal subsampling point is still given by $q_s = a_{s} + \frac{\Delta_s}{2}$. So, we use the relationship between $q_s$ and $\Delta_s$ as defined in Table~\ref{tab:coefs} for all types of systems we consider.

\begin{table}
    \caption{\label{tab:coefs} Optimal choice of the subsampling point $q_s$ in terms of the thickness $\Delta_s$ of each radial interval for 3D, 2D and 1D semiconductors. The inner radius of each interval is denoted by $a_s \equiv \sum_{i=1}^{s-1} \Delta_s$. We also compare the first optimal point $q_1$ for different dimensionality.}
    \begin{ruledtabular}

    \begin{tabular}{ccc}
	D & $q_s$ & $q_1$ \\
    \hline
	3 & $\sqrt{a_s^2 + a_s\Delta_s + \Delta_s^2/3}$ & $0.577\Delta_1$ \\
	2 & $a_{s} + \frac{\Delta_s}{2}$ & $0.500\Delta_1$ \\
	1 & $ {(a_{s}+\Delta_s) (1 + \Delta_s/a_{s})^{a_{s}/\Delta_s}}/{e}$ & $0.368\Delta_1$
    \end{tabular}
    \end{ruledtabular}

    
      
\end{table}


\bibliography{subsampling}
\bibliographystyle{apsrev4-1}

\end{document}